\pdfoutput=1
\documentclass{article}

\usepackage{epsf,graphicx}
\usepackage[hidelinks]{hyperref}
\hypersetup{pdftitle={Gliders on Aperiodic Monotilings: Cellular
Automata on the Hat and Spectre}, pdfauthor={James Edmond}}
\usepackage{amsmath}
\usepackage{color} 
\usepackage{amssymb,ComplexSystems}
\newcommand{\inert}[1]{\textcolor[gray]{0.5}{#1}}

\newif\ifpreprint
\preprinttrue
\makeatletter
\ifpreprint
  \imagefalse 
  \renewcommand{\@firstcopyrightfoot}{\hfil{\scriptsize Preprint.}\hfil}
  \def\@evenfoot{{\scriptsize Preprint}\hfil}
  \def\@oddfoot{\hfil{\scriptsize Preprint}}
  \def\@evenhead{\footnotesize{\sffamily\bfseries\thepage}\hfil\slshape Gliders on Aperiodic Monotilings}
  \def\@oddhead{\footnotesize{\slshape Gliders on Aperiodic Monotilings}\hfil{\sffamily\bfseries\thepage}}
\fi
\makeatother

\newenvironment{keywords}%
  {\par\vskip6pt\small\leftskip=2pc\noindent\textbf{Keywords:}\ }%
  {\par}

\begin{document}

\title{Gliders on Aperiodic Monotilings:\\
Cellular Automata on the Hat and Spectre}

\author{\authname{James Edmond}\\[2pt]
\authadd{London, United Kingdom}\\
\authadd{james@offlattice.org}}

\markboth{Complex Systems}
{Gliders on Aperiodic Monotilings}

\maketitle

\begin{abstract}
The hat and spectre monotiles, discovered in 2023, tile the plane only
aperiodically. Cellular automata are studied here on patches
generated by finite-state transducers, so that every experiment
regenerates deterministically from a small record. Within
edge-adjacency semi-totalistic rules, exhaustive and evolutionary
searches find only mortal travelers: gliders are absent. Guided by a
reproduction of the known Penrose-tiling glider, the rule space is
extended to vertex neighborhoods and to priority-table rules whose
non-quiescent states are visible to neighbors. Evolutionary search
then discovers gliders on both monotilings; tracked by a sliding
window that regenerates the patch along the flight, they travel one
million rings at constant speed and heading. All headings are
quantized, to millidegrees, onto a six-spoke compass --- the fast axes
of the tiling's graph metric. An ablation shows both rule-space
extensions are individually necessary. All results replay exactly in
an accompanying interactive essay.
\end{abstract}


\section{Introduction}
\label{sec:intro}

An aperiodic monotile is a single shape whose copies tile the plane,
but never periodically. Whether such a shape exists---the
\textit{einstein} problem---remained open for over half a century,
until 2023: the \textit{hat}, a 13-sided polykite, tiles the plane
only aperiodically \cite{hat}, and the \textit{spectre}, found weeks
later, does so without reflected copies \cite{spectre}. The two
tilings have been studied intensively as geometry; here they are
substrates for dynamics.

\begin{figure}
\centerline{\includegraphics[width=25.5pc]{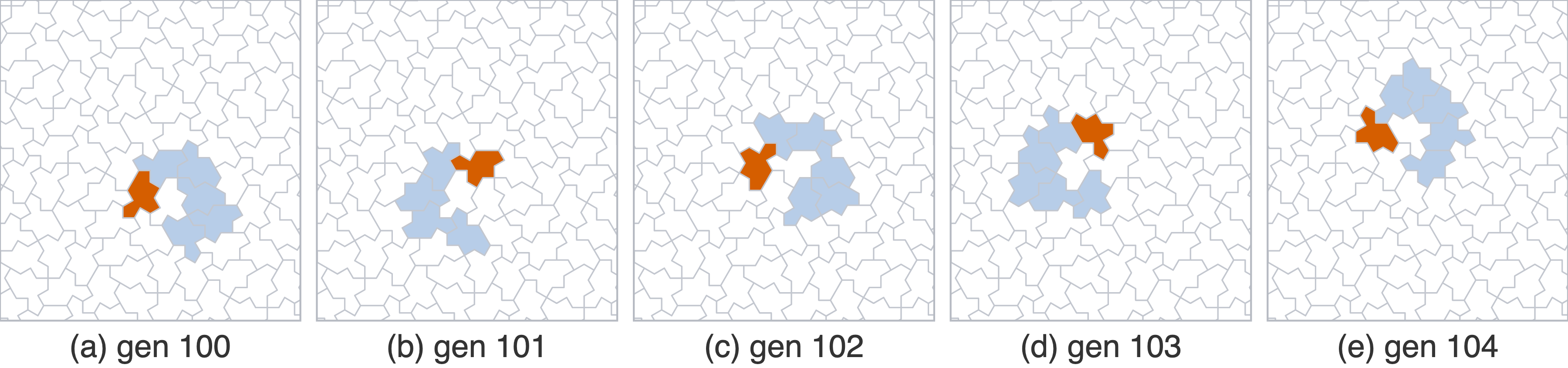}}
\caption{Five consecutive generations of a glider on the spectre
tiling. Gray outlines show a patch of the spectre tiling --- a
tiling with no translational period; colored tiles are the
non-quiescent cells of a four-state
cellular automaton---one \textit{head} cell in state 1 (vermilion)
and a wake of state-3 \textit{wing} cells (pale blue); state 2,
which this rule never enters, is green where later automata use it.
State colors darken toward state 1, so the ordering survives
grayscale reproduction. The panels are consecutive generations (100--104 of the
committed record), so each transition can be checked by hand against
the rule table; the object, one of a pair launched by a two-cell
seed, holds a fixed heading and advances one ring of the tiling every
two generations, its handful of cells re-formed at every step on
tiles it has never visited. The rule and seed are given in
Section \ref{sec:atlas}; the record replays live in the companion
essay.}
\label{fig:spectreglider}
\end{figure}

This paper asks whether the monotilings support \textit{gliders}:
bounded patterns that travel indefinitely. In Conway's Game of Life
\cite{gardner} the glider is the elementary carrier of information,
and everything Life is famous for---guns, logic gates, universal
construction \cite{winningways}---rests on it. On a periodic lattice, gliders are cheap
in a precise sense: translation symmetry guarantees that a pattern
that works once works forever, because after one period it faces an
identical neighborhood. An aperiodic tiling revokes exactly this
guarantee --- not because local environments are unrepeatable (every
finite patch of an aperiodic tiling recurs infinitely often), but
because no translation maps the tiling to itself: recurrence comes
with no schedule, and surviving one environment implies nothing about
the next. A moving pattern cannot translate; it must be re-formed,
step by step, along a sequence of local environments that never
becomes periodic. Figure \ref{fig:spectreglider} shows an object
that does so.

The question has a history on the Penrose tilings \cite{penrose}.
Owens and Stepney
studied Game of Life rules on the kite-and-dart and rhomb tilings and
mapped their still lifes and oscillators \cite{owens-stepney}; no
mobile pattern was found. The first glider on any aperiodic tiling
was exhibited in 2012 \cite{goucher}: a four-state automaton on the
generalized Moore neighborhood whose two-cell seed travels the rhomb
tiling indefinitely, overturning a conjecture---backed by a modest
cash prize---that no such object could exist \cite{trevorrow}. Its
immortality is geometric. The rhomb tiling carries \textit{de Bruijn
ribbons} \cite{debruijn}, straight bands of rhombi crossed edge to
parallel edge, and
the glider rides them; because rhombi are parallelograms, progress
along a ribbon is monotone and the path can never close. By a
different route, Bailey and Lindsey constructed automata on
multigrid-derived Penrose tilings that are isomorphic to Life
itself, importing its glider wholesale rather than finding one
native to the tiling \cite{baileylindsey}. On the hat tiling itself,
rules have been explored interactively since the year of its discovery
in Trevorrow's \textit{HatLife} \cite{hatlife}, including a
hand-designed rule whose \textit{worms} travel across its patch; we
learned of that work only after the first version of this paper was
posted, and a note at the end of Section \ref{sec:discussion} sets it
beside the results below. The monotilings have no analogue of the rhomb
tiling's straight rails. Goucher's mechanism therefore does not
transfer, and the question becomes: can directed transport exist
without rails?

Our answer is yes, and the paper is organized around how such a claim
can be tested. Every experiment is a compact record---
tiling family, root address, radius, neighborhood, rule, and seed---
from which patch and run regenerate deterministically via the
substitution-trans\-ducer machinery of Tatham \cite{tatham};
classification is exact, with periods detected as exact recurrences
of the full patch state. Searches are evolutionary, and their fitness
functions carry structural controls that were each installed after a
search exploited the objective's previous form (Section
\ref{sec:methods}). Above all, the pipeline was validated before any
monotiling claim was trusted: pointed at the Penrose rhomb tiling,
where ground truth exists, it must and does reproduce the published
glider---a reproduction that also surfaced a misprint in the
published transition table---and, unaided, rediscovers gliders there
at a measurable search cost (Section \ref{sec:penrose}). The same
compiled engine that ran every search runs in the browser: a
companion interactive essay replays each committed record live, so
the evidence for every figure in this paper can be watched, not
merely cited (Section \ref{sec:repro}).

The contributions are as follows.

\begin{enumerate}
\item Machinery for cellular automata on the hat and spectre tilings in
which every run is an exactly replayable record and every patch is
regenerated from its address rather than shipped as geometry.
\item A scoped negative result: under edge adjacency and
Generations-style rules (dying states invisible to neighbors), the
monotilings support only mortal travelers---exhaustively so for two
states, and a conclusion that survives a thirteenfold increase in
search budget.
\item A reproduction of the published Penrose glider with a corrected
transition table (the misprint is confirmed against the paper's
prose, the reference implementation, and reproduction), together with
blind rediscovery of gliders on that tiling, including one twice the
published speed.
\item Gliders on both the hat and the spectre: population-bounded
travelers flown one million rings at constant speed and heading in a
sliding window, igniting from 71--89\% of arbitrary two-tile seeds,
with a first interaction repertoire (annihilation, nondestructive
deflection, and one observed lane capture).
\item A compass law, measured exact: every mobile object travels on
a fan of directions quantized to \(60^\circ\)
(\(36^\circ\) on Penrose P3) with one fitted offset per substrate;
on the monotilings all six spokes carry measured lanes, and the fan
directions are the cusp axes of the tiling's graph metric---the
Penrose rails emerge as a special case.
\item An ablation showing the two rule-space extensions (the vertex
neighborhood, and neighbor-visible states with conjunctive
conditions) are each individually necessary, and a transferable
anti-gaming methodology in which every parasite bred by a fitness
function became a structural control.
\end{enumerate}

Section \ref{sec:substrate} describes the substrate and engine;
Section \ref{sec:methods} the rule families, the search, and the
verification standard; Section \ref{sec:penrose} the Penrose
validation. Section \ref{sec:gliders} presents the monotiling
results: the gliders and their verification, the compass law,
interactions, ignition genericity, and the ablation. Section
\ref{sec:atlas} is an atlas of the individual objects --- for each,
the complete rule, seed, portraits, and measured invariants,
everything needed to recreate and recognize it. Section
\ref{sec:discussion} discusses mechanism and prospects, and Section
\ref{sec:repro} records the reproducibility artifacts.

\section{Substrate and Engine}
\label{sec:substrate}

\subsection{Patches as graphs}

A cellular automaton on a tiling takes the tiles themselves as cells.
Following the combinatorial approach of Tatham \cite{tatham}, a
\textit{patch} is identified by a triple (tiling family, root tile
address, radius): starting from a canonical address for the root tile,
a breadth-first walk over the family's neighbor transducer enumerates
every tile within the given graph distance, producing an adjacency
graph in compressed sparse row form together with per-cell metadata
(tile class, chirality, address, and graph distance from the root).
The class records the tile's role in the substitution hierarchy ---
which slot of which first-level supertile it occupies, counting the
transducer's context variants of the supertiles as distinct
\cite{tatham} --- and takes one of finitely many values: 17 on the
hat tiling, 10 on the spectre.
The dynamics never see coordinates. Polygon geometry is computed
separately, by exact arithmetic in a cyclotomic ring, and is used only
for rendering and for the vertex adjacency defined below; because each
tile is placed once via its breadth-first tree edge, every non-tree
graph edge independently checks that two placements derived along
different paths still abut, validating the transducer against the
geometry.

Patch generation is deterministic: the same triple always yields the
same graph, with the same cell numbering, and the cells of a
radius-\(r\) patch form an index-for-index prefix of the cells of any
larger patch on the same root. Every experiment in this paper is
therefore stored as a small \textit{record}---family, root, radius,
neighborhood, rule, and initial cell states---from which the run
regenerates exactly, in the search code and in the browser alike.

One pitfall is recorded here because correctness depends on it: an
eventually periodic address whose repeating cycle touches its
supertiles' boundary describes a \textit{cone} rather than the
plane, and the transducer silently glues the seam, corrupting
adjacency near the seed. Root addresses are therefore chosen with
\textit{eventually interior} cycles, and patch generation flags any
seam-crossing cell; every run in this paper has an empty flag set.

\subsection{Two neighborhoods}

Two adjacency relations generalize naturally from grid automata to
polygonal tilings: tiles sharing an edge (the generalized von
Neumann neighborhood \cite{vonneumann} --- four neighbors on the
square grid) and tiles sharing
at least one boundary vertex (the Moore neighborhood \cite{moore},
as generalized by Owens and Stepney \cite{owens-stepney} --- eight
on the square grid, the neighborhood of Life). Edge adjacency is what the neighbor transducer
provides natively. Vertex adjacency is computed at patch-generation
time by collecting each tile's placement vertices in exact arithmetic
and linking tiles that share a vertex value; equality is exact, so no
tolerance parameter is involved, and the result is again a plain
adjacency graph.

Neighborhood sizes are variable, which shapes the available rule
families (Section \ref{sec:methods}). Under edge adjacency, interior
hat-tiling cells have four to six neighbors (the reflected tiles,
anti-hats, always have exactly four), spectre cells have four to
seven, and Penrose rhombs and kites have exactly four. Under vertex
adjacency, hat and spectre cells have six or seven neighbors, Penrose
P2 cells eight to ten, and P3 cells seven to eleven.

Activity is measured in \textit{rings}: the graph distance of a cell
from the root under edge adjacency, which is also the metric of patch
construction. A single vertex-adjacency step can cross more than one
ring, because the tiles around a shared vertex form an edge-connected
fan; the maximum ring difference across any vertex edge, the
\textit{fan margin}, is measured on each patch directly (two rings on
the monotilings and P2, three on P3). Cells within the fan margin of
the patch boundary have incomplete neighborhoods, and classification
treats them as boundary.

\subsection{Exact classification}

The engine performs synchronous, double-buffered updates on the fixed
graph, with a dead boundary. Runs are classified exactly. Cycle
detection uses Brent's algorithm \cite{brent} over full state snapshots, so a
reported period is an exact global recurrence, not a hash collision;
in particular a traveling pattern cannot be misreported as periodic,
because on an aperiodic substrate a mover never returns the whole
patch to a previous state. A run ends in one of five outcomes: died;
periodic, with exact period; reached the patch boundary; exceeded a
population cap; or still active at the horizon. Reaching the boundary
both terminates faithful simulation (the infinite tiling is no longer
being simulated) and serves as the explosion filter, since
space-filling rules reach it in time proportional to the radius.

\subsection{Flight-scale simulation: the sliding window}
\label{sec:slide}

A monolithic patch has cost quadratic in its radius --- near two
million tiles at radius 768, the practical ceiling --- so verification
by patch growth alone cannot follow a glider far. Because a patch is
only an address and a radius, the window can instead follow the
object. A flight is simulated on a small window (radius 24 suffices
for the gliders here); whenever the active set nears the window edge,
the entire nonzero state is extracted, a fresh window is generated
rooted at the leading cell's canonical address, the extracted cells
are placed into it by address, and the run continues. An invariant
asserted at every generation --- no active cell within the fan margin
plus two rings of the window edge --- guarantees that every
neighborhood the automaton ever reads is complete, so by induction the
composite run is exactly the run on the infinite tiling. The
construction is also validated directly: sliding runs reproduce
monolithic runs exactly at every sampled generation.

Geometry survives the hops. Consecutive windows share thousands of
tiles; matching them by address determines the rigid transform between
the two window frames exactly --- the rotation must lie on the
tiling's finite orientation set, and this is asserted --- and
composing the transforms yields the object's global position and
heading from launch, to floating-point precision, over arbitrarily
many hops. Memory use is constant in flight distance, per-hop cost
grows only logarithmically (the depth of the window's address), and
every hop emits a complete record, independently replayable like any
other. This is the instrument behind the million-ring measurements of
Section \ref{sec:gliders}; the companion essay animates the mechanism
--- the substrate generated just ahead of the glider as it flies.

\section{Rule Families, Search, and Verification}
\label{sec:methods}

\subsection{Rule families}

Variable neighborhood sizes force \textit{semi-totalistic} rules:
transitions may depend on the number of neighbors in given states, but
not on their arrangement --- the class the Penrose-tiling literature
calls \textit{outer totalistic} \cite{owens-stepney,goucher}. Two
families are used.

The first is the \textit{Generations} family. A rule has \(k\) states:
0 is quiescent, 1 is alive, and states \(2, \ldots, k-1\) are dying
phases that age deterministically toward death. Only neighbors in
state 1 are counted; a dead cell becomes alive if its live-neighbor count lies
in a birth set \(B\), a live cell remains alive if its count lies in a
survival set \(S\) and otherwise begins dying. For \(k = 2\) this is
exactly the Life-like family. Rules may additionally be
\textit{stratified} by tile class, giving each class its own
\((B, S)\) table --- 17 tables on the hat, 10 on the spectre.

The second family, \textit{priority-table rules}, removes two
restrictions at once. A rule is an ordered list of rows; each row
specifies an own-state (or a wildcard), a conjunction of per-state
neighbor-count thresholds (for example, \(n_1 \geq 1\) and
\(n_3 \geq 2\), where \(n_i\) counts neighbors in state \(i\)), and a
next state. The first matching row fires; if none matches, the cell
becomes quiescent. Two properties distinguish this family from
Generations: non-quiescent states other than 1 are visible to
neighbors, and conditions may require several signals jointly. The
family strictly contains the Generations rules, since any exact
count can be encoded by descending thresholds. Its significance for
gliders is developed in Section \ref{sec:penrose}.

\subsection{Search}

Two-state edge-adjacency rule spaces are small enough to enumerate:
about \(10^4\) to \(10^5\) rules per family after excluding birth on
zero neighbors, each evaluated on random soups and on small seeds and
classified exactly. Roughly three quarters of the hat tiling's
two-state rule space is explosive.

Larger spaces are searched by a genetic algorithm. For Generations
rules the genome is one \((B, S)\) table per tile class; for
priority-table rules it is the row list itself (at most eight rows of
at most two conditions), mutated by row insertion, deletion,
reordering, and field edits, and recombined by one-point row splicing.
Populations are initialized at random; no known rule is planted.
Each genome is evaluated from a fixed bank of 13 two-cell and
one-ball seeds at the patch root, neutral with respect to state roles,
and scored by
\begin{equation}
f \;=\;
\begin{cases}
0 & \text{if } P_{\max} > C, \\[2pt]
d \cdot P_{\mathrm{inner}} / P_{\max} & \text{otherwise},
\end{cases}
\label{fitness}
\end{equation}
where \(d\) is the farthest ring reached by any state change,
\(P_{\max}\) is the peak population, \(C\) is a population cap, and
\(P_{\mathrm{inner}}\) is the peak population before activity first
reached half the boundary distance. The ratio in
equation~(\ref{fitness}) is a growth-slope probe: a bounded traveler
keeps \(P_{\mathrm{inner}} \approx P_{\max}\), while a pattern that
grows as it travels is discounted.

\subsection{Verification, and the parasites that shaped it}

A bounded pattern that reaches the boundary of one patch proves only
that it traveled at least that radius. The glider criterion used
throughout is therefore \textit{cross-radius verification}: the
candidate must reach the boundary at every radius in a geometric
family (48, 96, 192, 384 rings here), with flat peak population at
every scale. The spacing is multiplicative because tightly spaced
radii can mask slow growth.

Each term of the fitness function, and each verification control,
exists because a search exploited its absence. Reward for reaching the
boundary let mortal travelers pose as escapees on any patch smaller
than their range (hence raw distance, and verification only across
radii). The population cap alone bred ever-thinner growers that fit
under it (hence the growth-slope ratio). Most strikingly, on
edge-adjacency patches the table-rule search discovered
\textit{boundary sniffing}: a rule such as ``any cell with
\(n_0 \geq 3\) becomes quiescent; otherwise state 1'' ignites nothing
that travels, but detects the patch boundary itself, because corner
cells of a finite patch have fewer neighbors than any interior cell.
The exploit scales with the patch and therefore passes cross-radius
verification---the only parasite found that does. It is excluded by a
\textit{causality filter}: activity must remain inside the light cone
of the seed, that is, the farthest changed ring may not exceed
\(d_0 + g \cdot m\), with \(d_0\) the seed's extent, \(g\) the
elapsed generations, and \(m\) the fan margin.
Boundary sniffing is structurally impossible under vertex adjacency,
where the minimum degree exceeds the threshold range, and none of the
gliders reported below relies on it; the general lesson, that every
objective breeds its parasite and every parasite must become a
structural control, is one of the paper's methodological conclusions.

\section{Validation on Penrose Tilings}
\label{sec:penrose}

Before trusting any search on substrates where we had no ground
truth, the pipeline was pointed at the one aperiodic tiling with a
peer-reviewed glider. Gliders on Penrose tilings were conjectured impossible
by Trevorrow \cite{trevorrow} and found by Goucher \cite{goucher}: a
four-state outer-totalistic automaton on the generalized Moore
neighborhood whose head-and-tail seed travels along the de Bruijn
ribbons of the P3 rhomb tiling---straight bands of rhombs crossed
edge to parallel edge, whose parallelogram geometry forbids the path
from ever closing. The validation had two levels: can the machinery
\textit{recognize} the known glider, and can the search
\textit{rediscover} gliders unaided at feasible cost?

\subsection{Detection, and an erratum}

Encoded exactly as printed, the published transition table cannot
glide: no row of the table causes a change to state 1, so a head can
never be reborn, and the seed dies within four generations in every
orientation. Yet the paper's own prose describes exactly that event
--- ``a new head is reborn ahead of the original one'' --- and only
the second row can implement it, since its conditions (a ground cell
beside the head and at least two of its wings) single out the cell
just ahead of the glider. Its next state must therefore be 1, not
the printed 3; the reference implementation distributed with the
simulator \textit{Ready} \cite{ready} uses 1. Table
\ref{goucher-table} shows the corrected automaton; with it, the
glider reproduces immediately.

\begin{table}
\centerline{\small\begin{tabular}{|c|c|c|}
\hline
Current state & Neighbor condition & Next state \\
\hline
0 & \(n_1 \geq 1\) and \(n_2 \geq 1\) & 3 \\
\hline
0 & \(n_1 \geq 1\) and \(n_3 \geq 2\) & 1 \\
\hline
1 & \(n_3 \geq 1\) & 2 \\
\hline
1 & always & 1 \\
\hline
2 & always & 3 \\
\hline
any & otherwise & 0 \\
\hline
\end{tabular}}
\caption{The corrected Penrose glider automaton of \cite{goucher}
(states: 0 ground, 1 head, 2 tail, 3 wing; first matching row fires).
The published table misprints the second row's next state as 3, under
which state 1 is unreachable and the automaton supports no glider; the
value 1 is confirmed by the paper's prose, by the reference
implementation \cite{ready}, and by reproduction.}
\label{goucher-table}
\end{table}

With the corrected rule, a head-and-tail pair seeded across any shared
edge of any tile class launches a glider (40 of 40 seedings tested),
which crosses patches of radius 24 to 120 at a constant two
generations per ring with peak population ten, flat at every radius.
On the P2 kite-and-dart tiling the same rule instead produces
\textit{loopers}, closed orbits whose exact periods---20, 40, and, on
larger patches, 200---match the published values; 40 and 200 open
the published decagonal sequence \(40, 200, 1240, \ldots\), whose
successive periods obey \(P_{n+2} = 5 P_{n+1} + 6 P_n\).
Exactness here is meaningful: periods come from Brent
recurrence of the full patch state, not from local pattern matching.

\subsection{Rediscovery, and a calibration}

The second level asks whether the unmodified search finds gliders
where gliders exist. At modest budget (population 96, 300
generations) it does not: blind runs plateau on mortal travelers and
oscillators at a quarter of the attainable fitness. Seeding one run
with the known glider confirmed that the objective ranks a true
glider strictly above everything the blind searches had found; the
optimizer, not the fitness landscape, was the limit. At a sevenfold
larger budget (population 256, 800 generations, about
\(1.15 \times 10^5\) genome evaluations), blind search crossed the
needle: four of six independent runs produced cross-radius-verified
gliders.

None of them is Goucher's. One is a \textit{phoenix} glider whose
rule gives states 1 and 2 no survival rows at all---every live cell
dies each generation and the pattern is reborn ahead of itself---and
which travels at one ring per generation, twice the speed of the
published glider, verified flat to radius 384; its seed emits two
gliders separated by \(142.9^\circ \approx 4 \times 36^\circ\).
Its automaton is Table \ref{tab:phoenixrule}, printed verbatim from
the evolved genome in the convention of Section \ref{sec:atlas}. The
remaining finds are \textit{wanderers}: cohesive, ash-free patterns
of 17 to 32 cells whose displacement grows as \(t^{\alpha}\) with
\(\alpha \approx 0.6\)--\(0.85\), roughly thirty times farther than
diffusion over the measured span but short of ballistic, realized as
straight runs along pentagrid directions punctuated by stochastic
reorientation. Two further full-fitness champions were correctly
rejected by verification, one locking into an exact period-145 orbit
after 128 rings, one dying at 48. The rediscovery also calibrates
search cost: the searches behind the negative result
(Section \ref{sec:ablation}) ran nearly an order of magnitude below the
threshold at which gliders become findable on a substrate known to
contain them.

\begin{table}
\centerline{\small\begin{tabular}{|c|c|c|}
\hline
Current state & Neighbor condition & Next state \\
\hline
3 & \(n_2 \geq 1\) & 1 \\
\hline
0 & \(n_2 \geq 3\) & 0 \\
\hline
0 & \(n_3 \geq 1\) and \(n_2 \geq 1\) & 3 \\
\hline
\inert{0} & \inert{\(n_3 \geq 1\) and \(n_2 \geq 1\)} & \inert{3} \\
\hline
0 & \(n_3 \geq 2\) and \(n_3 \geq 2\) & 1 \\
\hline
\inert{0} & \inert{\(n_3 \geq 2\) and \(n_3 \geq 2\)} & \inert{2} \\
\hline
0 & \(n_3 \geq 1\) and \(n_0 \geq 3\) & 2 \\
\hline
\inert{0} & \inert{\(n_3 \geq 1\) and \(n_0 \geq 3\)} & \inert{2} \\
\hline
any & otherwise & 0 \\
\hline
\end{tabular}}
\caption{The phoenix glider's automaton (first matching row fires;
the final row is the default; grayed rows are inert in the sense of
Section \ref{sec:atlas}). Seed: state 2 on the root thick rhomb,
state 3 on a thin neighbor. States 1 and 2 have no survival rows ---
every live cell dies each generation and the anatomy is reborn ahead
of itself, one ring per generation. Row two, dead-to-dead, is
nonetheless essential: it suppresses births beside the exhaust.}
\label{tab:phoenixrule}
\end{table}

\section{Gliders on the Monotilings}
\label{sec:gliders}

\subsection{Discovery}

The pipeline of Section \ref{sec:penrose} was pointed at the
monotilings unchanged: vertex neighborhood (measured fan margin two),
four-state priority-table rules, random initialization, the same
neutral seed bank, the same fitness (equation (\ref{fitness})) with
population cap 64, searched at radius 48 with population 256 for 800
generations --- the budget at which gliders are routinely findable on
the Penrose rhomb tiling. Three independent runs were made on each
family.

\begin{figure}
\centerline{\includegraphics[width=21pc]{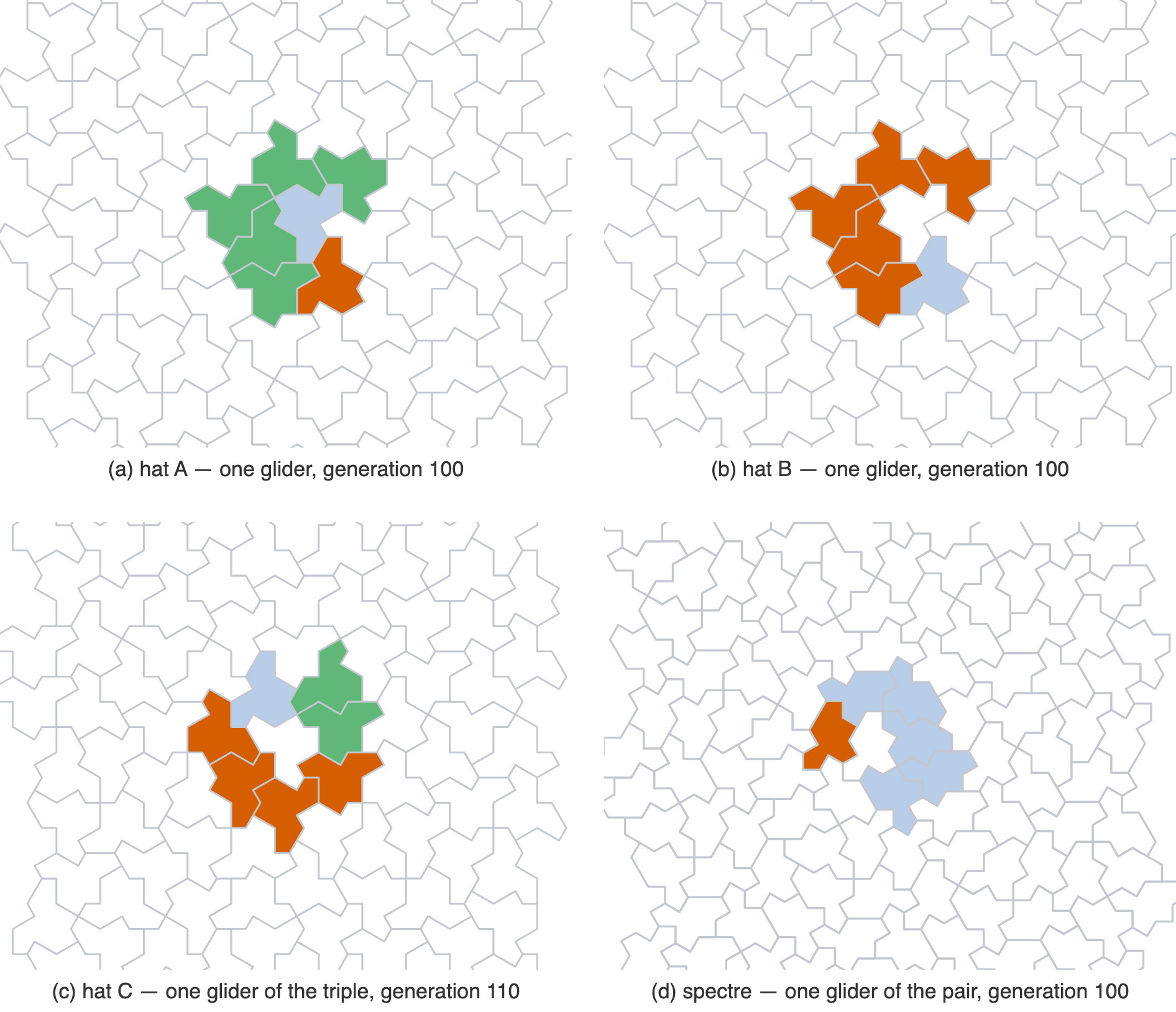}}
\caption{The four monotile gliders, one object each at a common
scale --- the glider body is the rule's invariant, shared by every
seeding, while launch multiplicities and spread patterns are seed
artifacts. (a) Hat glider A, a compact object of five to seven
cells, at generation 100; (b) hat glider B at the same generation
--- visibly a different anatomy (its active cells alternate sides
across the movement axis), yet it travels with the identical clock
and heading as A; (c) one glider of hat C's triple at generation
110, shortly after the wandering intermediate decayed; (d) one
glider of the spectre pair at generation 100 --- the most minimal
anatomy found, a single bright head and its wing wake. Colors as in
Figure \ref{fig:spectreglider}.}
\label{fig:portraits}
\end{figure}

Four of the six runs produced cross-radius-verified gliders: three on
the hat and one on the spectre. Where the Penrose rediscovery had
required about \(1.15 \times 10^5\) genome evaluations to cross its
needle, the first hat run reached full fitness at genetic generation
five --- roughly 1,300 evaluations. The two
remaining runs were rejected by verification, and both rejections are
instructive; they are discussed below. Figure \ref{fig:portraits}
shows the four objects. Their complete automata and seeds, each with
a space-time rendering and a full-lifetime trail, are collected in
the atlas of Section \ref{sec:atlas}; the rule behind
Figure \ref{fig:spectreglider} appears there as
Table \ref{tab:spectrerule}.

\subsection{Verification and the shared clock}

\begin{figure}
\centerline{\includegraphics[width=25.5pc]{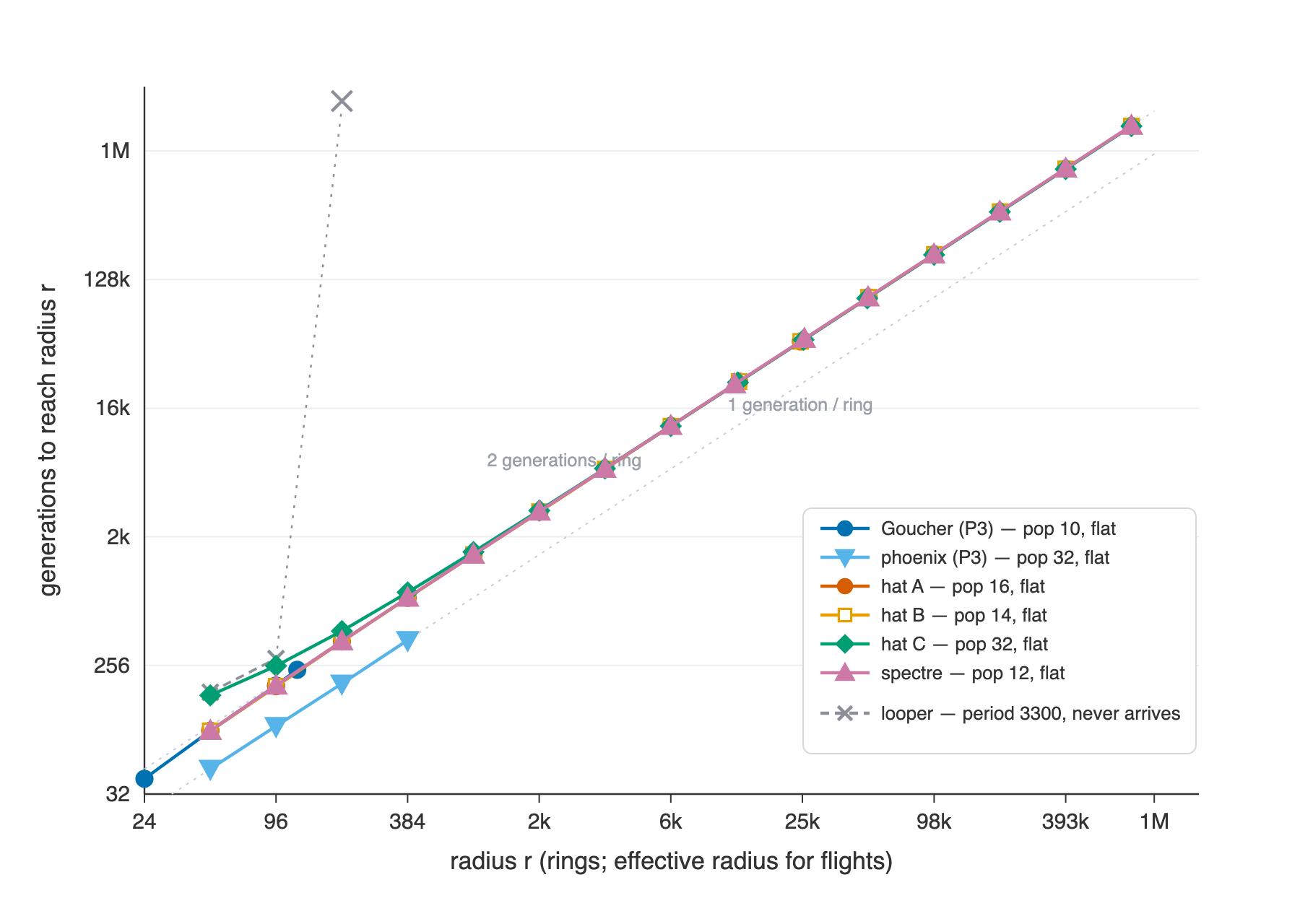}}
\caption{Generations to reach graph distance \(r\) from the seed,
versus \(r\), log--log. Points up to radius 768 are monolithic
boundary arrivals; beyond, each line continues as a sliding-window
flight plotted against effective radius, to one million rings. A
glider is ballistic: slope one. All four monotile gliders share the
clock of two generations per ring and fuse into a single bundle
across more than three decades; the phoenix runs at one generation
per ring. Arrivals are noisy at launch and converge onto the clock
within a few hundred generations. The gray dashed curve is the
rejected looper: at radius 192 both its wanderers are captured
by closed loops (periods 60 and 550; full-state recurrence 3,300)
and it never arrives (Section \ref{sec:atlas}).}
\label{fig:verification}
\end{figure}

Figure \ref{fig:verification} shows the verification data. Each
glider was replayed on monolithic patches of radius 48, 96, 192, 384,
and 768 --- a sixteenfold span, the largest a patch of about two
million tiles --- and reached the boundary at every radius with its
peak population exactly flat: 16, 14, 32, and 12 cells respectively,
at every scale. Each was then flown by sliding window
(Section \ref{sec:slide}) to one million rings --- roughly 1,300
times the monolithic ceiling --- with the same flat population and
the same clock throughout, every hop an independently replayable
record. The terrain crossed is never revisited, and the sequence of
local environments along the lane, though its members recur, never
becomes periodic.

The slopes carry a regularity that nothing selected for: the fitness
functions reward directed travel with flat population and say nothing
about speed. Yet five unrelated rules --- Goucher's, designed by hand
fourteen years earlier, and hat A, hat B, hat C (after its transient), and
the spectre pair, evolved here --- share a single clock of two
generations per ring across three substrates, with the phoenix at
exactly one.
We read the common value as the natural pace of relay transport:
these objects do not translate but are re-formed ahead of themselves,
and one ring of progress appears to cost one excitation--promotion
cycle of the relay. The clock is predictive out of sample: fitted at
radii up to 384, it puts the radius-768 arrivals near generation
1,536, and the measured values are 1,533--1,534.

What these verifications do and do not establish must be stated
plainly. On a periodic lattice, gliderhood has a finite certificate:
a pattern that recurs translated is immortal by symmetry. Goucher's
glider has a geometric substitute (monotone progress along a ribbon).
Neither is available here, and no finite replay can prove immortality
on an aperiodic substrate; \textit{glider} in this paper is the
operational term defined by the verification standard, not a proven
property. The in-family cautionary case is the spectre looper of
Figure \ref{fig:verification}: its pair of wanderers traveled with
near-flat population (29 rising to 31), one reaching ring 189 before
being captured by a closed loop at generation 992 --- so late
transitions are real, and the million-ring flights stand more than
three decades beyond the deepest capture observed.
Two measured properties argue specifically against a hidden loop or
late trap: loopers turn, while these gliders hold their settled
headings to within hundredths of a degree over the entire flight
(Section \ref{sec:compass}); and their arrival laws are predictive
out-of-sample, where an approach to a loop would bend them. A path
toward genuine proof is discussed in Section \ref{sec:discussion}.

\subsection{Launch, interactions, and ignition}
\label{sec:interactions}

\begin{figure}
\centerline{\includegraphics[width=25.5pc]{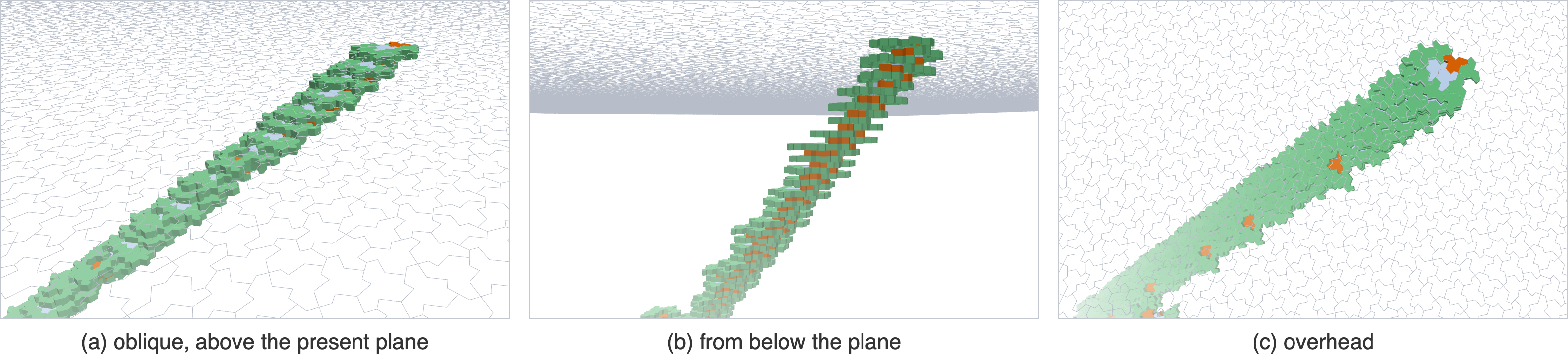}}
\caption{Hat glider A's history as space-time geometry: each
generation extrudes the non-quiescent tiles as prisms, time upward,
older layers fading over 120 generations; the wireframe is the tiling
at the current slice. (a) Oblique view from above the present plane;
(b) from below --- the state-1 footprints tick at exactly two
generations per ring, the relay clock as visible rungs, with the
launch transient the kink at the tube's foot; (c) overhead. Speed is
slope, the generation-42 collision is a vertex where one tube ends on
the other, and the tube's lean never leaves its compass direction.
Rendered by the same engine that runs every figure live in the
companion essay.}
\label{fig:worldtube}
\end{figure}

Launch multiplicity is a property of the seed, not the rule: the
heading sweep of Section \ref{sec:compass} later found clean
single-glider two-cell seeds on every hat spoke, and on most spectre
lanes. The champion seeds, as it happens, all launch more than one
object. Hat A's emits two gliders that travel together, braided a few
tiles apart, until one strikes the other's wake at generation 42 and
dies, leaving the survivor to run indefinitely. Hat B launches
the same way. Hat C ignites through a bound intermediate: a wanderer
that meanders near the origin before decaying into two, then three
gliders; its 62-generation arrival offset is the cost of this phase.
The spectre seed emits a clean pair separating onto lanes
\(120^\circ\) apart.
Figure \ref{fig:worldtube} renders hat A's launch as space-time
geometry.

Ignition does not depend on the discovery site. Seeding each rule's
state pair across every shared edge of one interior anchor per tile
class, in both orders, launches a verified-signature traveler from
71--89\% of seedings (hat A 167 of 188; hat B 165 of 188; spectre 81
of 114). Goucher's rule scores 100\% on this test, but
unsurprisingly so: a head-and-tail pair is that glider's own small
phase, so the seed is already a glider mid-cycle, and every P3 edge
lies on a ribbon that supports the next step. The monotile seeds
match no settled configuration --- the gliders fly at populations of
six and above --- so their percentages measure something stronger:
the reach of the gliding attractor from arbitrary two-cell states.
Strikingly, not one
of roughly 490 seedings of those three rules died or froze ---
everything travels. Hat C ignites less often (45 of 188), consistent
with its wanderer-mediated launch; visual review resolved most of
its remaining seedings as multi-glider showers rather than growth,
so the percentages are lower bounds. Beyond the launch annihilation,
pairs at other sites interact \textit{nondestructively}: course
changes for one or both gliders, always settling at mutual
separations of \(60^\circ\), \(120^\circ\), or \(180^\circ\). The
interaction repertoire --- synthesis by pair launch, annihilation by
wake contact, deflection onto compass-commensurate headings --- is
raw material for any future collision engineering.

\subsection{The compass law}
\label{sec:compass}

\begin{figure}
\centerline{\includegraphics[width=25.5pc]{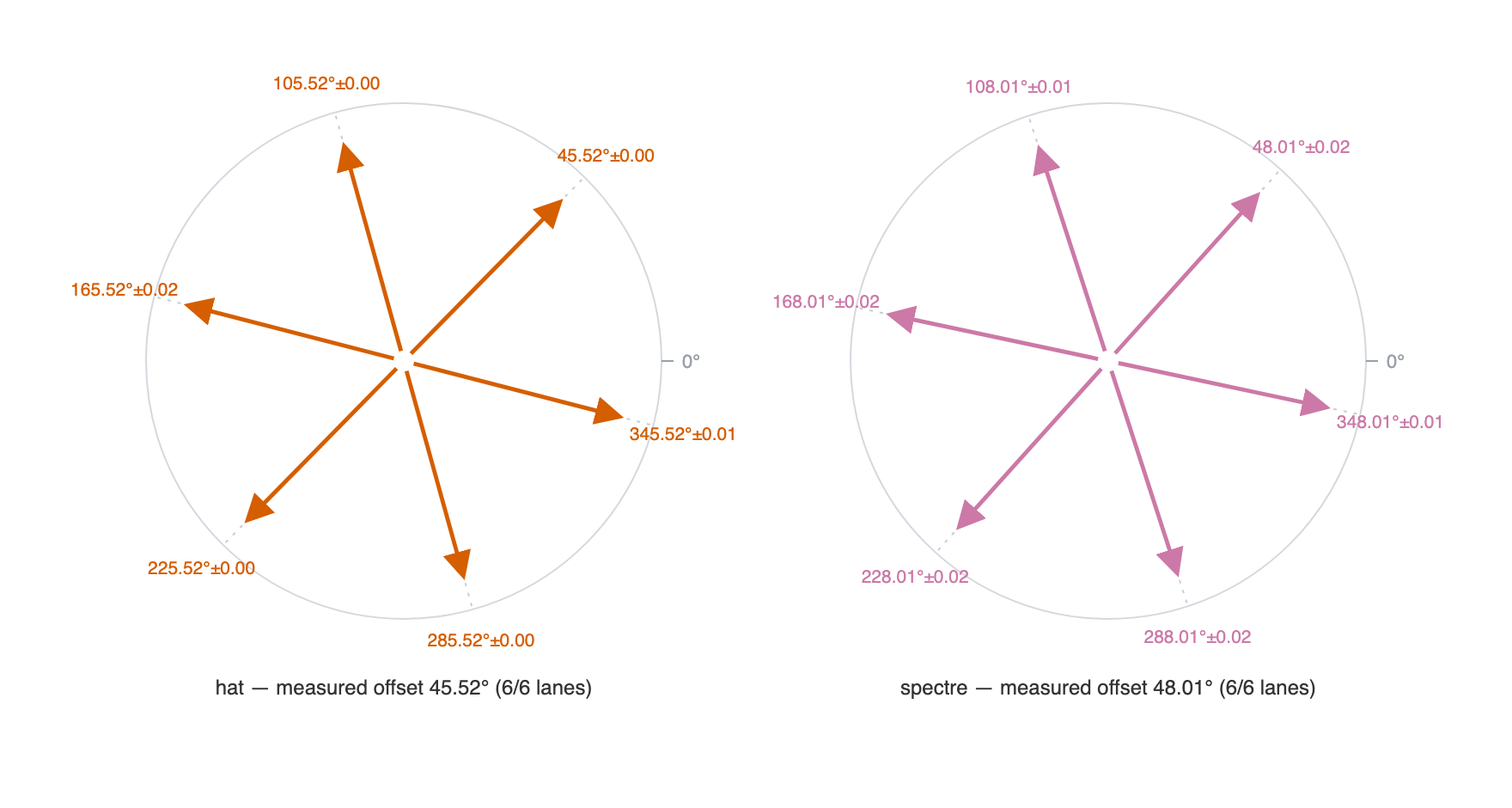}}
\caption{Flight-measured lane headings on the hat and spectre
(arrows; settled heading with standard error, from the
sliding-window flights). Every one of each substrate's six fan
spokes carries a measured glider lane, and all twelve headings agree
with a single per-substrate fan offset --- hat \(45.523^\circ\),
spectre \(48.014^\circ\) --- within standard errors of
\(0.001^\circ\)--\(0.024^\circ\); lane separations are
parameter-free multiples of \(60^\circ\). Wanderers are
omitted: their run segments lock to the fan, their net drift --- a
sum over reorientations --- does not.}
\label{fig:compass}
\end{figure}

Every mobile object in this paper travels along directions quantized
to its substrate's orientational order. On Penrose P3, measured by
windowed segment fits at patch scale, Goucher's glider locks to the
tenfold pentagrid fan to \(0.08^\circ\) mean windowed residual and
the phoenix pair to \(0.45^\circ\), separated by
\(142.9^\circ \approx 4 \cdot 36^\circ\); the coarser residuals
reflect the short baseline, not a weaker lock --- the Penrose objects
were not reflown at flight scale. On the monotilings the
flights measure the law at three more decades of baseline
(Figure \ref{fig:compass}): settled headings on all six spokes of
each substrate's sixfold fan --- twelve lanes, found by binning
generic-seed launches by heading --- every one agreeing with a single
per-substrate offset, hat \(45.523^\circ\) and spectre
\(48.014^\circ\) in the canonical frame of the root tile, within
standard errors of \(0.001^\circ\)--\(0.024^\circ\). Separations between lanes are
parameter-free: the hat triple's \(120.001^\circ \pm 0.011^\circ\),
the spectre pair's \(240.000^\circ \pm 0.009^\circ\) --- multiples
of \(60^\circ\) to within a hundredth of a degree. Even the Penrose
\textit{wanderers} --- cohesive, ash-free movers whose displacement
grows as \(t^{\alpha}\) with \(\alpha \approx 0.6\)--\(0.85\) ---
obey the law in segments: their straight runs lock to the pentagrid
fan at short windows, with stochastic reorientation between runs, so
the substrate sets the transport law while the rule sets the
reorientation rate.

Launch
transients --- start-point offsets and early reorientations --- decay
as \(1/r\), and the heading sweep caught one launch mid-capture: it
ran four window-hops near one spoke, reoriented by exactly
\(60^\circ\) in a single hop, and held the adjacent lane for the
remaining hundred thousand rings (the record replays in the
interactive essay). Lane-locking, on this evidence, is an attractor
process: the six spokes are attracting states of the relay, a
transient can include a whole lane switch, and no settled glider has
ever been observed to leave its lane.

\begin{figure}
\centerline{\includegraphics[width=25.5pc]{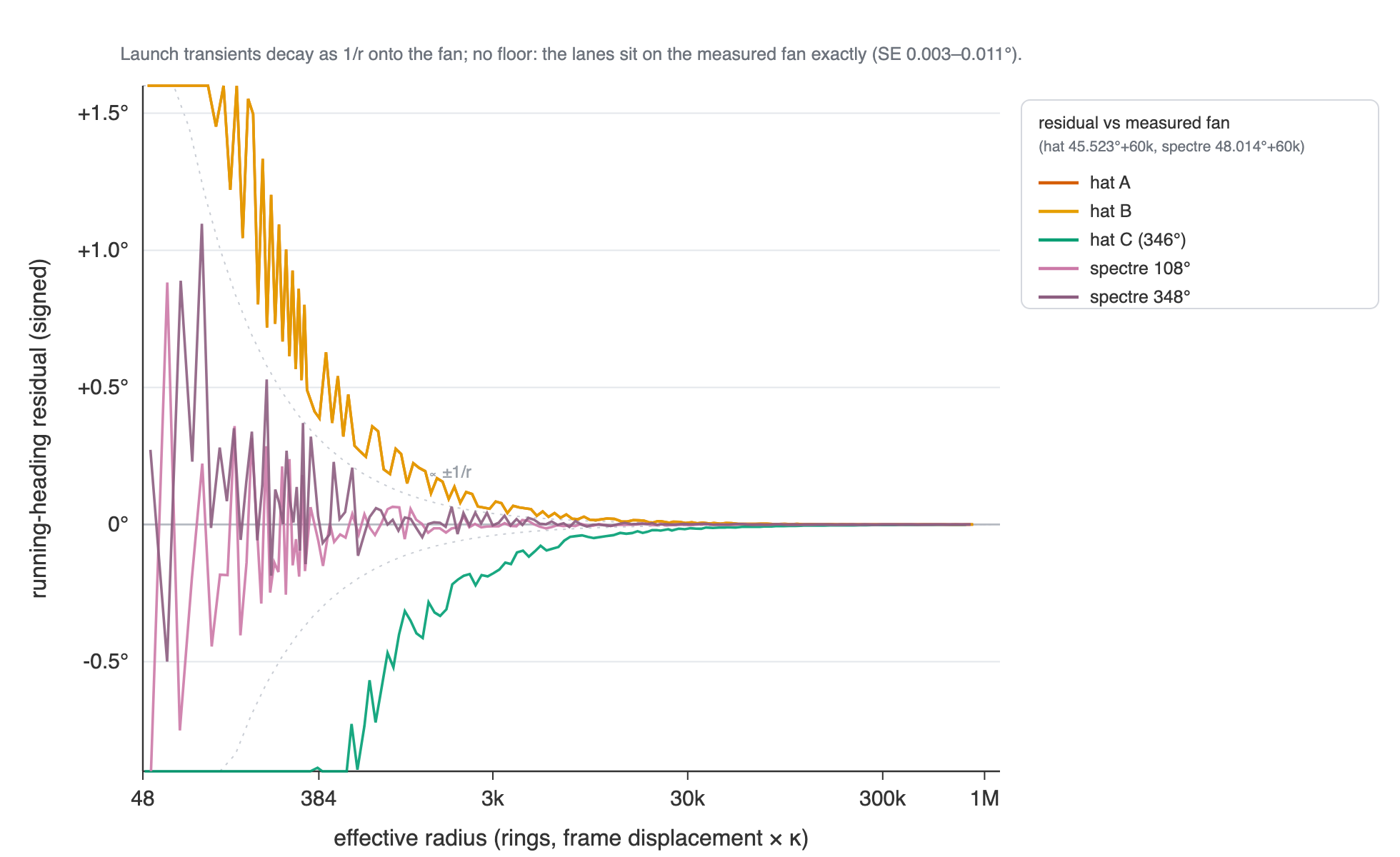}}
\caption{Signed residual of the running heading from the nearest
spoke of the measured fan, versus effective radius. Launch
transients decay as \(1/r\) (dotted envelopes) onto the fan;
beyond about thirty thousand rings every lane's residual is
statistically zero. Three hat gliders from independent genomes ---
two sharing a lane, the third \(120^\circ\) away --- and both
members of the spectre pair (\(240^\circ\) apart) land on the
same per-substrate fan. Hat A lies exactly beneath hat B; hat C's
larger transient is its wanderer-mediated launch.}
\label{fig:convergence}
\end{figure}

Figure \ref{fig:convergence} shows the approach to the fan.
Running headings converge as \(1/r\) --- the launch transient's
start-point offset averaging out --- and there is no residual floor:
beyond roughly thirty thousand rings every lane's residual is
statistically zero, with standard errors of
\(0.003^\circ\)--\(0.011^\circ\) at one million rings,
corresponding to about fifty tiles of transverse drift over the
entire flight. The lanes are real objects with
measurable widths: Goucher's ribbon-guided glider deviates less than
one tile from a straight line over a hundred tiles, while the
monotile gliders, with no rail beneath them, hold lanes two to four
tiles wide. Rails, on this evidence, were never the requirement for
directed transport --- the orientational order that every
substitution tiling carries suffices to steer, and the Penrose
ribbons are simply the case in which the lane is rigid.

The fan's phase --- why \(45.523^\circ\) --- is itself substrate
geometry. The graph-metric ball of each tiling is a rounded hexagon,
far from circular: Euclidean reach per ring varies with direction by
15\% on the hat and 12\% on the spectre, peaking in six sharp cusps.
Locating the cusps by flank intersection on a radius-384 patch places
all twelve, six per substrate, at the measured lane directions to
within \(0.15^\circ\), with the reach at each cusp matching the
flights' rings-per-length constant to a few parts per thousand. The
gliders travel the fast axes of their substrate's metric, at the
metric rate: the compass phase is a constant of the tiling, not of
the rule. (Chaotic growth fronts, by contrast, are circular to about
1\% on these same tilings --- growth is dynamically smoothed and
does not probe the cusps.)

\subsection{The ablation}
\label{sec:ablation}

Section \ref{sec:penrose} identified two extensions to the
originally searched space: the vertex neighborhood, and rules
whose non-quiescent states are visible to neighbors under
conjunctive conditions. To attribute the gliders, the full
two-by-two grid was searched at the calibrated budget --- two
independent runs on each family per cell, each cell searched by its
established machinery. Table \ref{tab:ablation} gives the outcome:
both extensions are individually necessary, and only their
combination suffices.

\begin{table}
\centerline{\small\begin{tabular}{|c|c|c|}
\hline
 & Edge adjacency & Vertex adjacency \\
\hline
Generations (\(k = 4\)) & mortal travelers, 43--58 rings & mortal
travelers, 45--58 rings \\
\hline
Priority tables & mortal travelers and oscillators, 31--47 rings &
\textbf{gliders (four of six runs)} \\
\hline
\end{tabular}}
\caption{The ablation grid at the calibrated search budget. The
edge-adjacency Generations cell reproduces the negative result at
thirteen times the budget of the searches that established it, so
that null was not underpowered. All three null cells share the same ceiling --- mortal
travelers of roughly 30--60 rings --- and only the combination of
vertex neighborhood and neighbor-visible rule states crosses from
mortal to verified gliding.}
\label{tab:ablation}
\end{table}

The null cells are informative beyond attribution. Generations rules
fail on both neighborhoods, even searched with per-class stratified
tables: their dying states are invisible to
neighbors, so the extra adjacency of the vertex neighborhood adds
nothing such a rule can read. On the square lattice the same
semantics are glider-prolific --- Brian's Brain is the \(k = 3\)
Generations rule \(B = \{2\}\) with empty survival \cite{toffoli}
--- so the
failure is a property of the pairing with aperiodic ground, not of
the family. Priority tables fail on edge adjacency even though the rule
family provably contains relay mechanisms. And all three null cells
plateau at the same 30--60-ring mortal ceiling that every search in
the deficient spaces runs into, exhaustive or evolutionary, on
either family --- suggesting that the ceiling is a property of the
rule spaces, not of the tilings, which the fourth cell confirms. The search in that cell
also produced a fourth fitness parasite, \textit{boundary sniffing}
(Section \ref{sec:methods}), whose structural fix, the causality
filter, is inherited by every search reported here.

\section{An Atlas of the Monotile Gliders}
\label{sec:atlas}

This section gives, for each of the four gliders and then for the
looper that the verification standard rejected, everything
needed to recreate and recognize it: the complete automaton, verbatim from
the evolved genome, with inert rows grayed out --- a row is grayed
when deleting it leaves every configuration of the verified replay
unchanged, so the black rows are the working rule and the gray are
genetic cruft (duplicated conjuncts, shadowed rows), reproduced
because the record is the rule; the final row of each table is the
default --- a close-up of
the object in flight, an oblique space-time rendering (conventions of
Figure \ref{fig:worldtube}: tiles extrude upward in time, 120-
generation fade, wireframe at the present), and the full lifetime of
the discovery run replayed at radius 384 --- an overhead map in which
every tile that was ever non-quiescent is tinted by the time of its
first activation, from pale sand at launch to vermilion at the
boundary, with the final generation in the state palette and the
patch edge drawn as the surrounding ring. Colors follow
Figure \ref{fig:spectreglider} throughout. Every panel regenerates
from the committed records.

A \textit{situation} is a snapshot of the glider as the rule sees
it: the non-quiescent cells together with the substrate
\textit{collar} within a fixed graph radius of them --- tile
classes, adjacencies, and vertex degrees --- canonically encoded, so
that two snapshots are equal exactly when their collared
neighborhoods are isomorphic, wherever on the tiling they occur. The
\textit{flying vocabulary} is the number of distinct situations an
object occupies over its verified flight; Section
\ref{sec:discussion} builds on the finiteness of this set.

\subsection{Hat glider A}

\begin{figure}
\centerline{\begin{minipage}[b]{11pc}\centerline{\includegraphics[width=11pc]{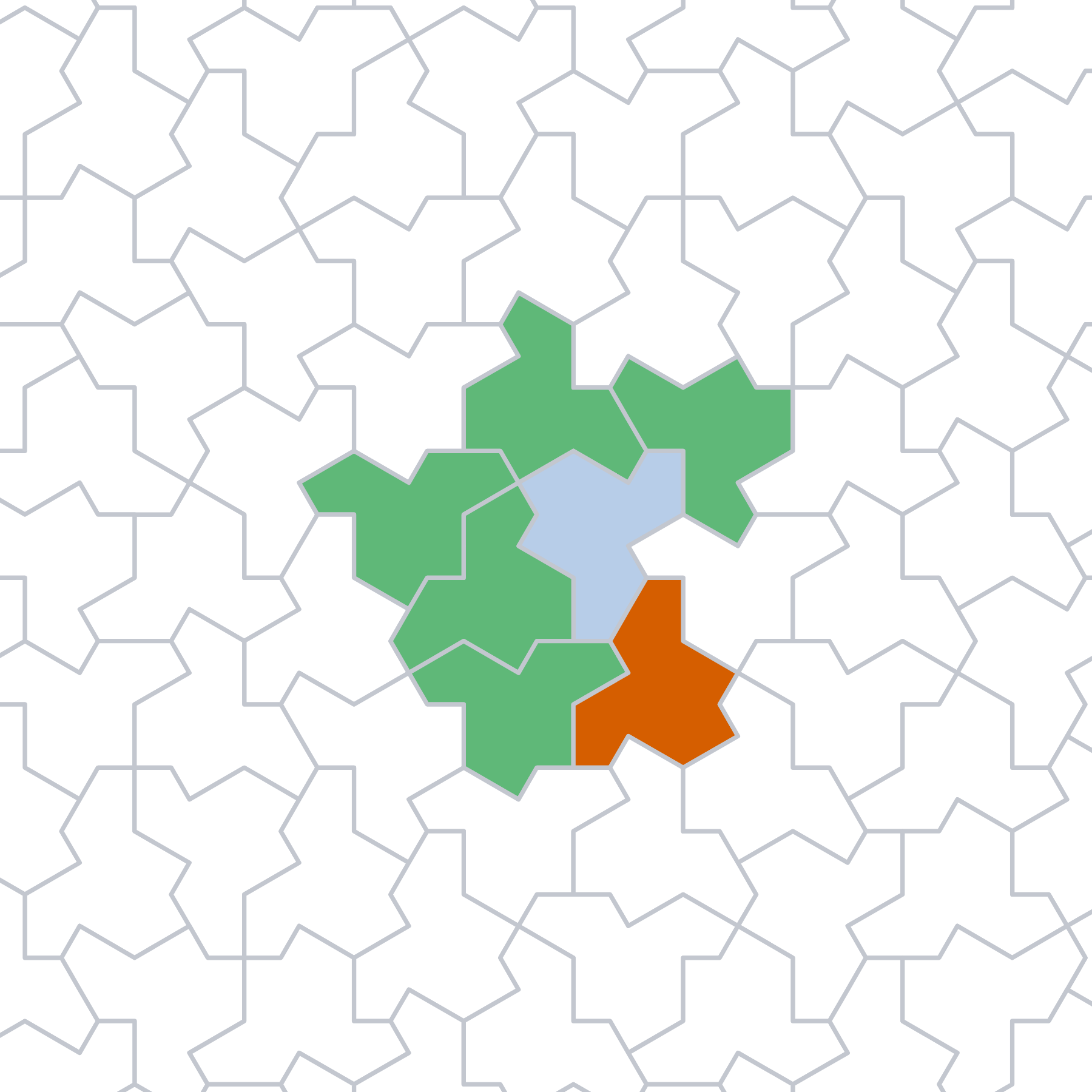}}\centerline{\footnotesize (a)}\end{minipage}\hskip 1pc
\begin{minipage}[b]{11pc}\centerline{\includegraphics[width=11pc]{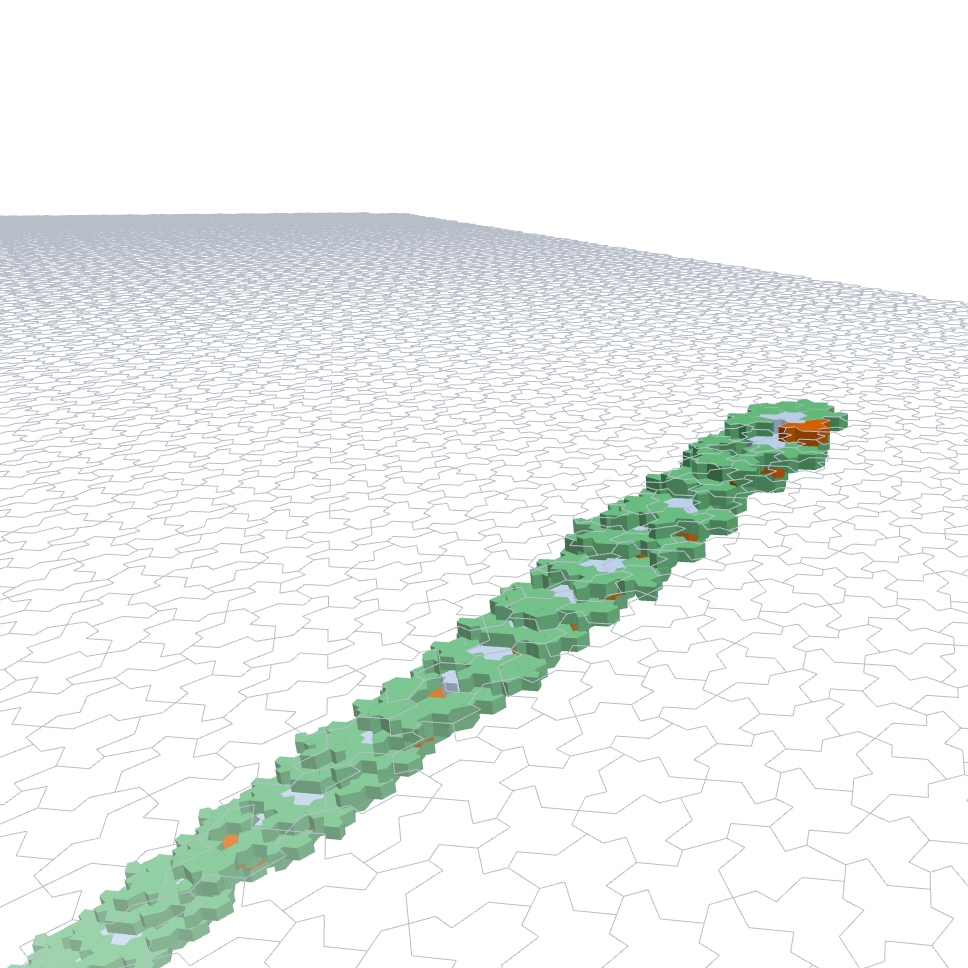}}\centerline{\footnotesize (b)}\end{minipage}}
\vskip 4pt
\centerline{\includegraphics[width=22pc]{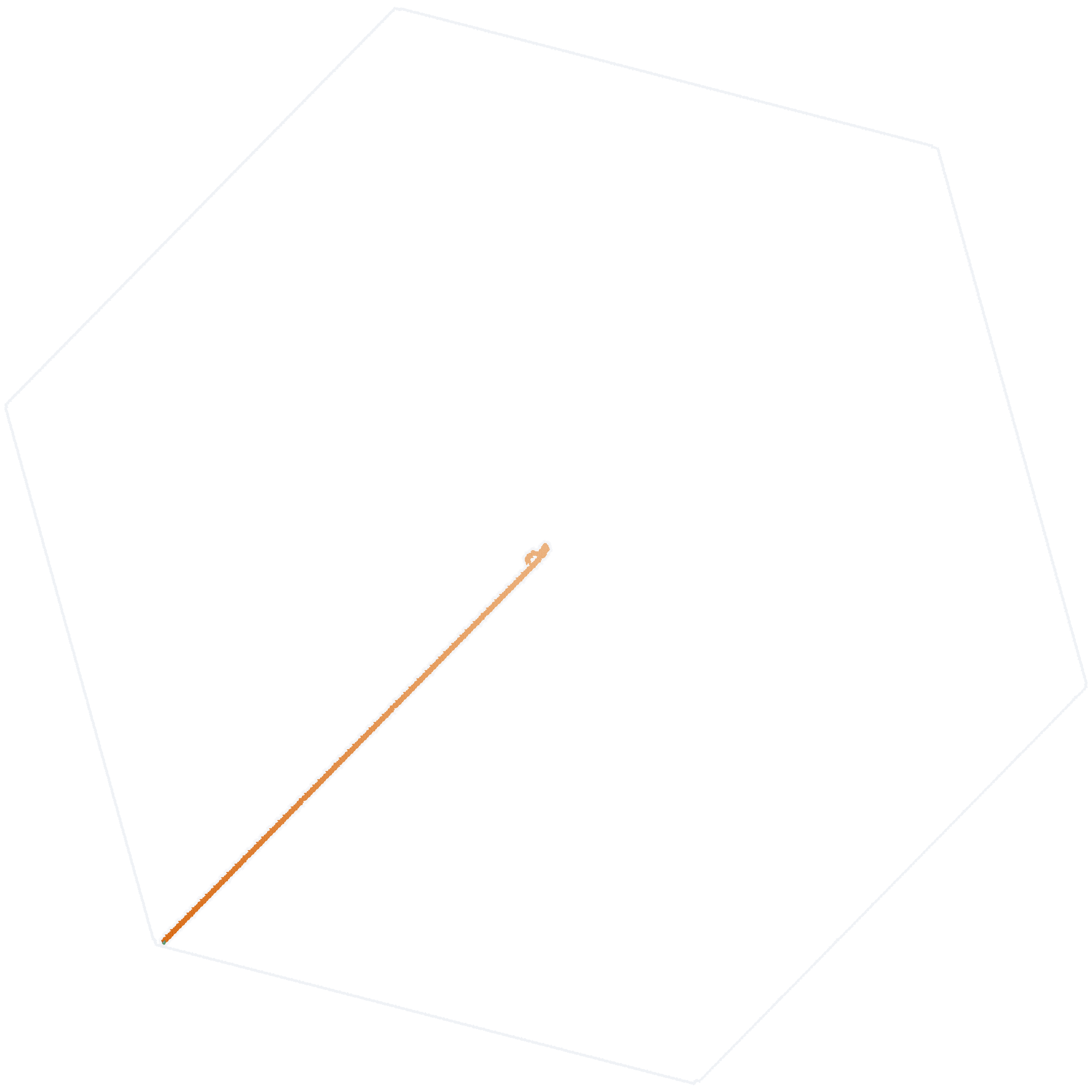}}
\centerline{\footnotesize (c)}
\caption{Hat glider A. (a) In flight at generation 100. (b)
Space-time worldtube, frozen at generation 150 mid-cruise. (c) Full
lifetime at radius 384: 759 generations, 1,942 tiles visited; the
pale curl at the ribbon's start is the two-glider launch and
generation-42 collision, and the ribbon meets the boundary at a
corner of the patch --- the compass law drawn by the object itself.}
\label{fig:atlas-s21}
\end{figure}

Record \texttt{hat-tableevolve-r48-s21}: peak population 16, two
generations per ring, settled heading \(225.523^\circ\) (a spoke
of the sixfold fan), launched by seeding states (1, 2) across one shared edge
at the patch root; 89\% of arbitrary two-tile seedings of this rule
launch travelers, and none die. Its flying vocabulary is 167
situations. The trail
(Figure \ref{fig:atlas-s21}c) records the launch: two gliders braid
from the seed until one dies on the other's wake at generation 42
--- the short stub at the pale end of the ribbon --- and the survivor
flies straight, meeting the boundary exactly at the southwest corner
of the patch.

\begin{table}
\centerline{\small\begin{tabular}{|c|c|c|}
\hline
Current state & Neighbor condition & Next state \\
\hline
0 & \(n_2 \geq 2\) & 0 \\
\hline
any & \(n_1 \geq 3\) and \(n_1 \geq 1\) & 3 \\
\hline
2 & \(n_1 \geq 1\) and \(n_0 \geq 1\) & 1 \\
\hline
1 & \(n_2 \geq 1\) and \(n_2 \geq 1\) & 3 \\
\hline
\inert{2} & \inert{always} & \inert{0} \\
\hline
0 & \(n_1 \geq 1\) & 2 \\
\hline
\inert{2} & \inert{\(n_3 \geq 2\)} & \inert{0} \\
\hline
\inert{2} & \inert{always} & \inert{0} \\
\hline
any & otherwise & 0 \\
\hline
\end{tabular}}
\caption{Hat glider A's automaton (first matching row fires). The
grayed rows are inert: all three send state 2 to quiescent, which is
the default fall-through.}
\label{tab:hatArule}
\end{table}

\subsection{Hat glider B}

\begin{figure}
\centerline{\begin{minipage}[b]{11pc}\centerline{\includegraphics[width=11pc]{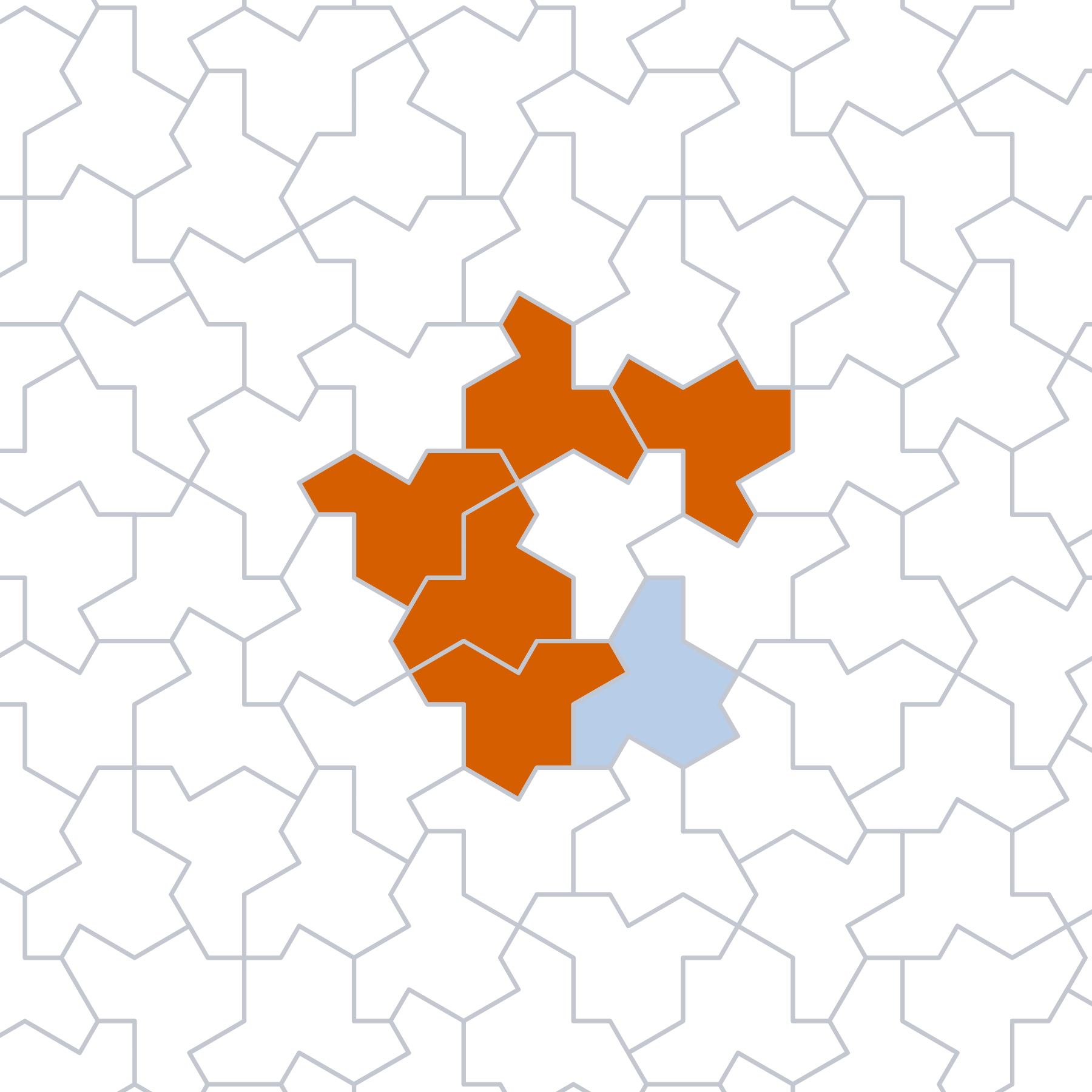}}\centerline{\footnotesize (a)}\end{minipage}\hskip 1pc
\begin{minipage}[b]{11pc}\centerline{\includegraphics[width=11pc]{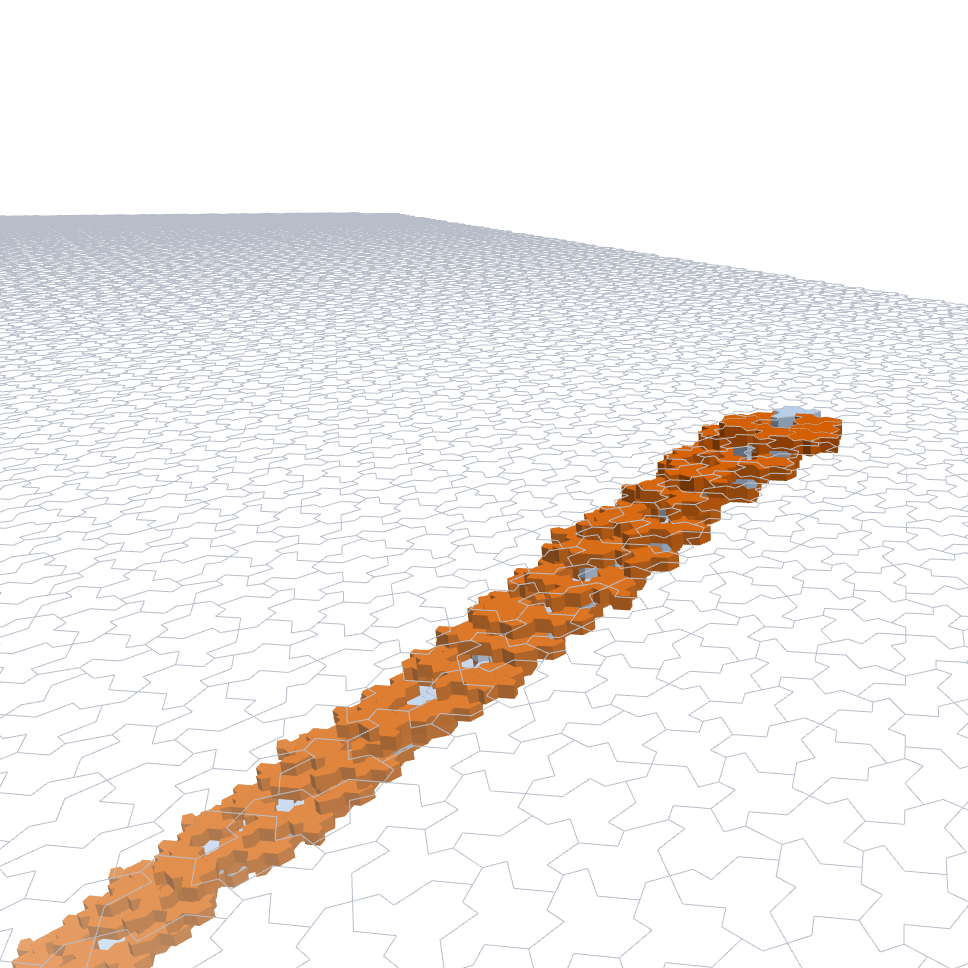}}\centerline{\footnotesize (b)}\end{minipage}}
\vskip 4pt
\centerline{\includegraphics[width=22pc]{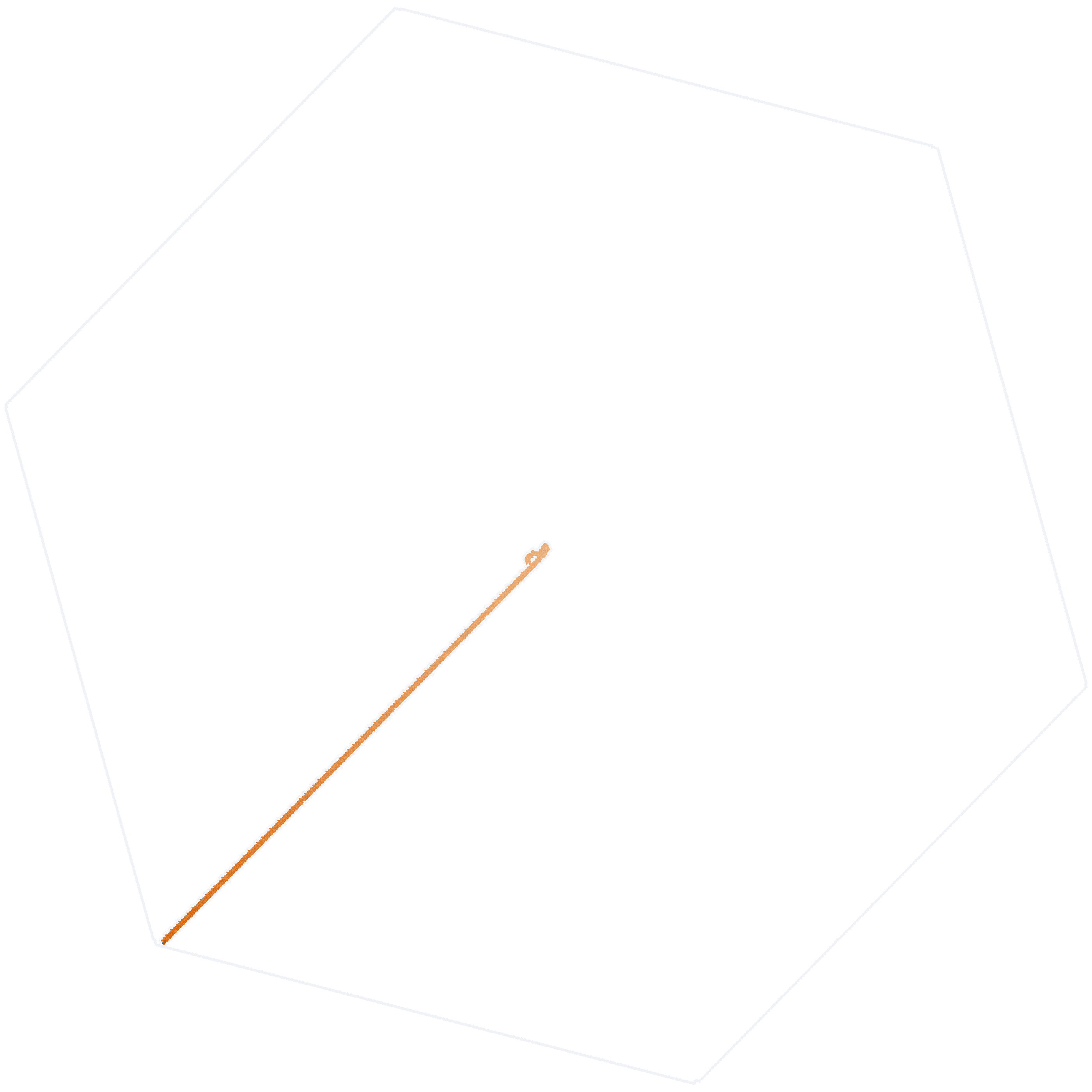}}
\centerline{\footnotesize (c)}
\caption{Hat glider B. (a) In flight at generation 100 --- the
state-1-heavy anatomy, visibly unlike A's. (b) Worldtube at
generation 150. (c) Full lifetime at radius 384: same launch stub,
same corner, same clock as A, from an unrelated genome.}
\label{fig:atlas-s22}
\end{figure}

Record \texttt{hat-tableevolve-r48-s22}: peak population 14, and the
same clock, heading, and windowed-heading sequence as hat A --- one
phenotype in two genomes --- but a different anatomy (four to five
state-1 cells alternating sides across the movement axis, where A
carries a state-2 body) and the tersest rule found: three working
rows.
Seed (3, 1); ignition 88\%, none die. Its launch also emits a braided
pair with one survivor, visible as the stub in
Figure \ref{fig:atlas-s22}c.

\begin{table}
\centerline{\small\begin{tabular}{|c|c|c|}
\hline
Current state & Neighbor condition & Next state \\
\hline
0 & \(n_1 \geq 2\) and \(n_0 \geq 1\) & 0 \\
\hline
\inert{0} & \inert{\(n_0 \geq 2\) and \(n_2 \geq 2\)} & \inert{1} \\
\hline
0 & \(n_0 \geq 2\) and \(n_3 \geq 1\) & 1 \\
\hline
1 & \(n_3 \geq 1\) & 3 \\
\hline
any & otherwise & 0 \\
\hline
\end{tabular}}
\caption{Hat glider B's automaton --- the tersest glider rule found.
State 1 has no survival row: a bright cell lasts one generation,
becoming a wing if a wing stands beside it and dying otherwise, so
the anatomy is reborn ahead at every step. The grayed row waits for
state 2, which neither the seed nor any row produces: it can never
fire.}
\label{tab:hatBrule}
\end{table}

\subsection{Hat glider C}

\begin{figure}
\centerline{\begin{minipage}[b]{11pc}\centerline{\includegraphics[width=11pc]{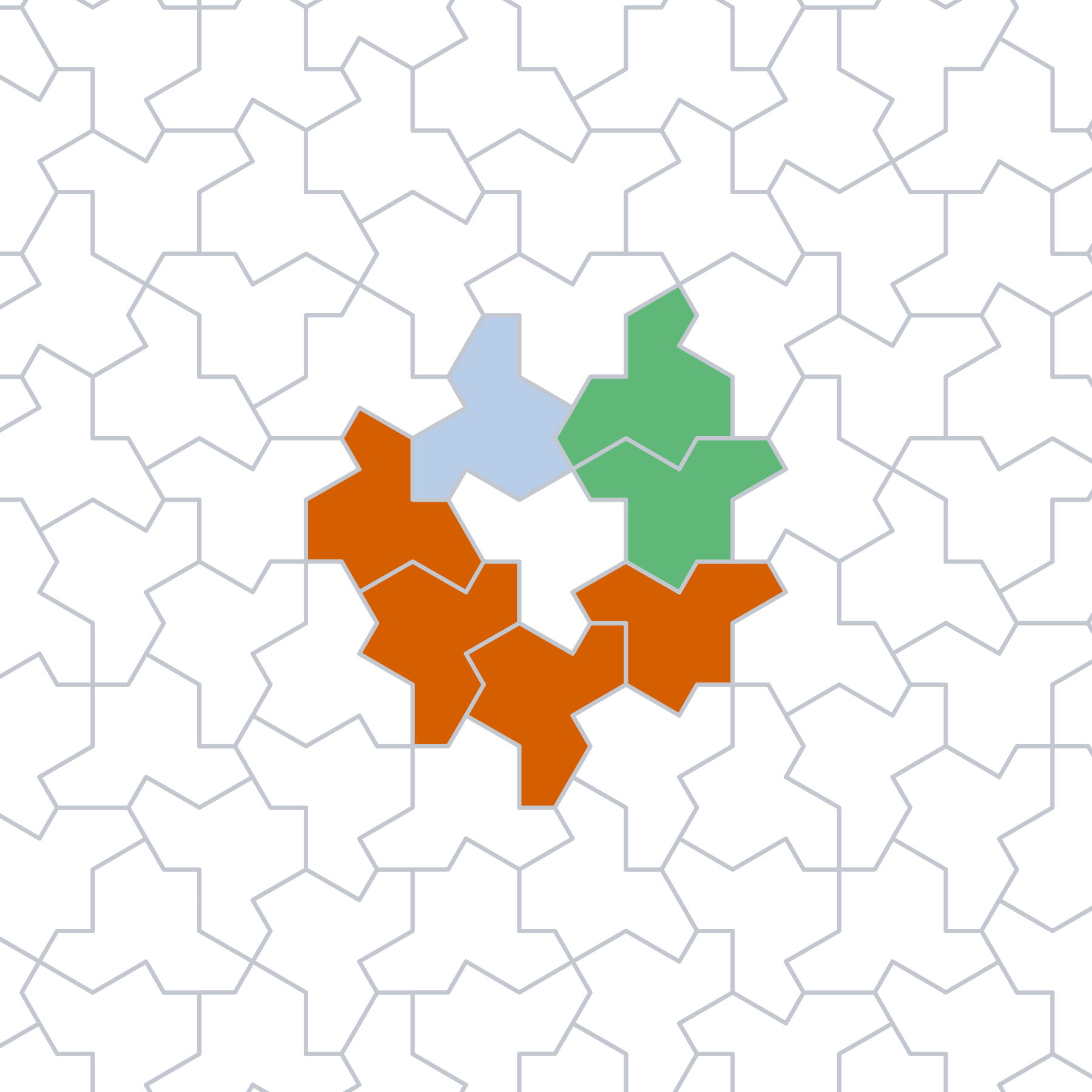}}\centerline{\footnotesize (a)}\end{minipage}\hskip 1pc
\begin{minipage}[b]{11pc}\centerline{\includegraphics[width=11pc]{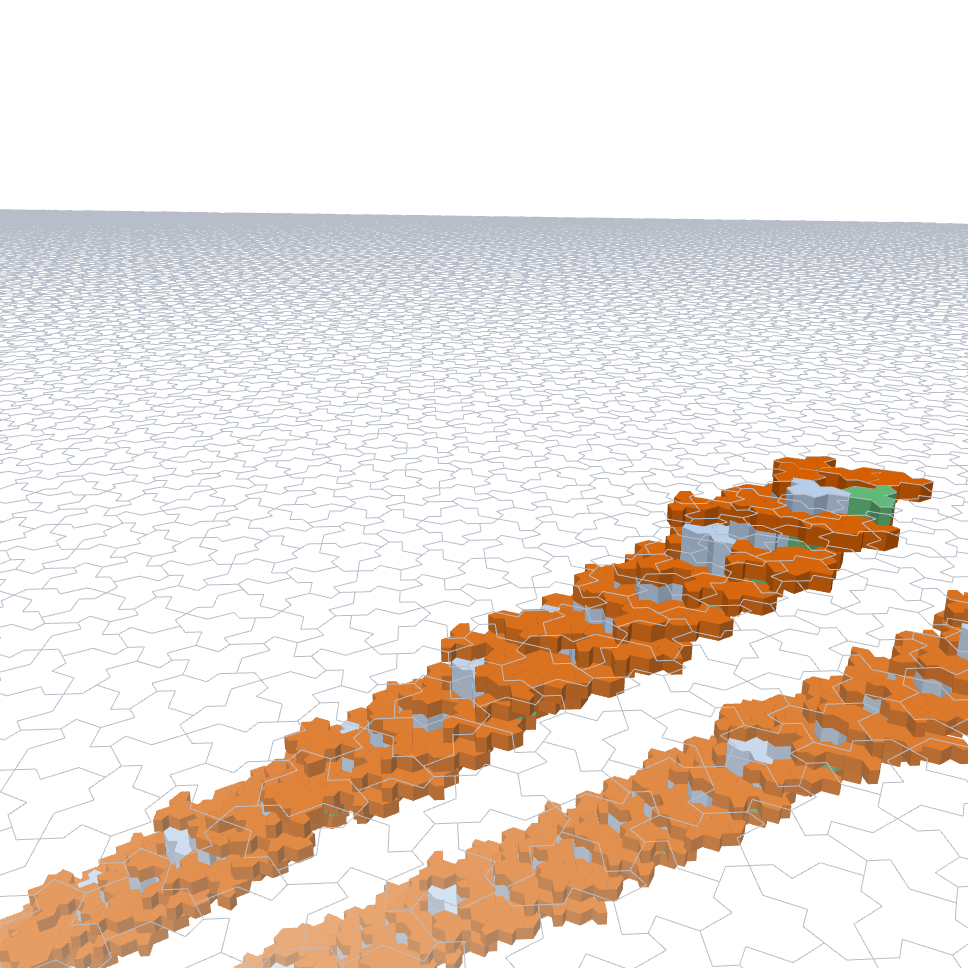}}\centerline{\footnotesize (b)}\end{minipage}}
\vskip 4pt
\centerline{\includegraphics[width=22pc]{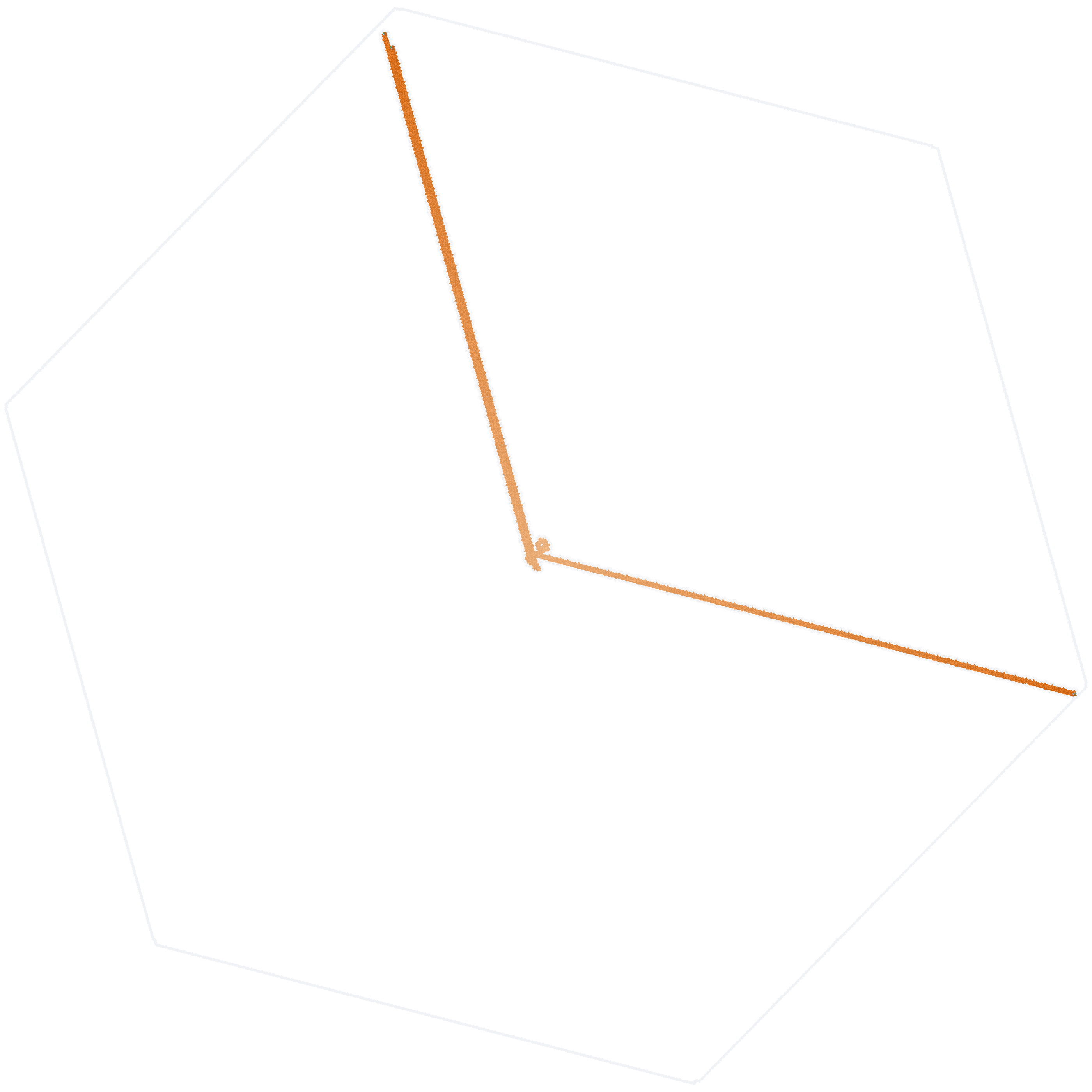}}
\centerline{\footnotesize (c)}
\caption{Hat glider C. (a) One glider of the \(105^\circ\) pair at
generation 110. (b) Worldtube of the parallel pair at generation 175
--- two tubes climbing side by side. (c) Full lifetime at radius 384:
830 generations, 5,167 tiles visited; the central knot is the
wandering intermediate, and the two ribbons --- the merged parallel
pair east-southeast and the lone glider north-northwest --- open
onto lanes \(120^\circ\) apart.}
\label{fig:atlas-s23}
\end{figure}

Record \texttt{hat-tableevolve-r48-s23}: the wanderer-mediated
triple. The seed (2, 1) ignites a bound intermediate that meanders
near the origin before decaying into three gliders --- a parallel
pair on the \(105^\circ\) lane and a third on \(345^\circ\), a
separation of \(2 \cdot 60^\circ\). Peak population 32 across all
three; arrival law \(\mathrm{gen}(r) = 62 + 2r\); ignition 24\%
(most other seedings launch multi-glider showers through the same
intermediate). The trail (Figure \ref{fig:atlas-s23}c) shows the
central knot of the wandering phase and the two ribbons leaving it.

\begin{table}
\centerline{\small\begin{tabular}{|c|c|c|}
\hline
Current state & Neighbor condition & Next state \\
\hline
any & \(n_2 \geq 1\) and \(n_1 \geq 3\) & 3 \\
\hline
1 & \(n_2 \geq 1\) & 2 \\
\hline
3 & \(n_3 \geq 3\) & 3 \\
\hline
0 & \(n_3 \geq 2\) & 1 \\
\hline
any & \(n_2 \geq 1\) & 1 \\
\hline
\inert{1} & \inert{\(n_0 \geq 3\) and \(n_2 \geq 2\)} & \inert{2} \\
\hline
\inert{0} & \inert{\(n_3 \geq 3\)} & \inert{3} \\
\hline
any & otherwise & 0 \\
\hline
\end{tabular}}
\caption{Hat glider C's automaton. The grayed rows are inert, each
shadowed by an earlier row: six by two (same outcome), seven by four,
which fires first on every cell seven matches.}
\label{tab:hatCrule}
\end{table}

\subsection{The spectre glider}

Record \texttt{spectre-tableevolve-r48-s33}: mechanically the most
minimal of the four --- a pure
three-state relay that never enters state 2 (Table
\ref{tab:spectrerule}: a head excites wings; a wing beside a head is
promoted to the next head; crowded wings die). Seed (1, 3); the pair
separates onto lanes \(120^\circ\) apart; peak population 12 for
both gliders;
the tightest compass lock measured (settled heading within
\(0.006^\circ\) of the fan over a million-ring flight); ignition
71\%, none die. Its flying vocabulary is 306
situations.

\begin{table}
\centerline{\small\begin{tabular}{|c|c|c|}
\hline
Current state & Neighbor condition & Next state \\
\hline
0 & \(n_0 \geq 3\) and \(n_1 \geq 1\) & 3 \\
\hline
\inert{0} & \inert{always} & \inert{0} \\
\hline
3 & \(n_3 \geq 3\) & 0 \\
\hline
\inert{0} & \inert{\(n_0 \geq 3\) and \(n_1 \geq 1\)} & \inert{0} \\
\hline
3 & \(n_1 \geq 1\) & 1 \\
\hline
\inert{0} & \inert{always} & \inert{0} \\
\hline
any & otherwise & 0 \\
\hline
\end{tabular}}
\caption{The spectre glider's automaton, the rule behind
Figure \ref{fig:spectreglider} (first matching row fires). The grayed
rows are inert: row two restates the quiescent default and rows four
and six sit in its shadow; the only live ground transition is row
one.}
\label{tab:spectrerule}
\end{table}

\begin{figure}
\centerline{\begin{minipage}[b]{11pc}\centerline{\includegraphics[width=11pc]{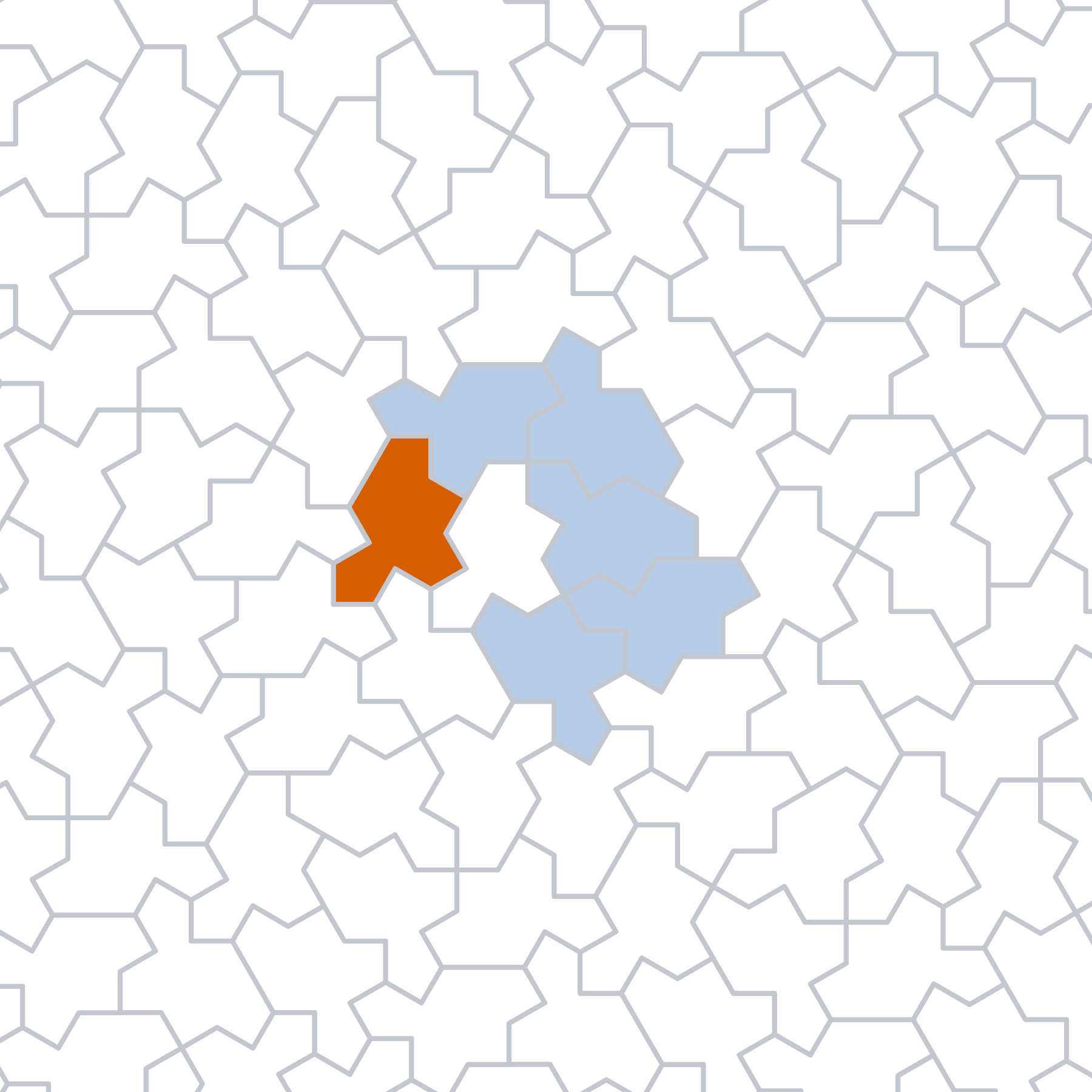}}\centerline{\footnotesize (a)}\end{minipage}\hskip 1pc
\begin{minipage}[b]{11pc}\centerline{\includegraphics[width=11pc]{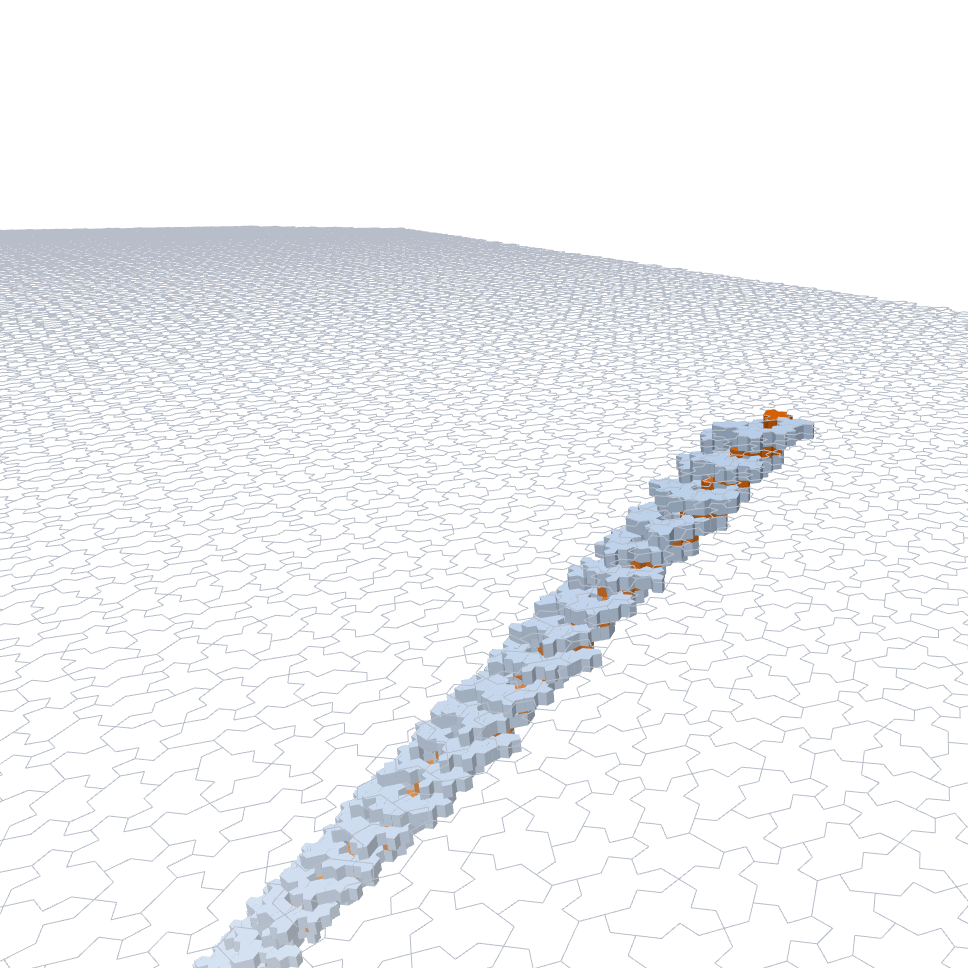}}\centerline{\footnotesize (b)}\end{minipage}}
\vskip 4pt
\centerline{\includegraphics[width=22pc]{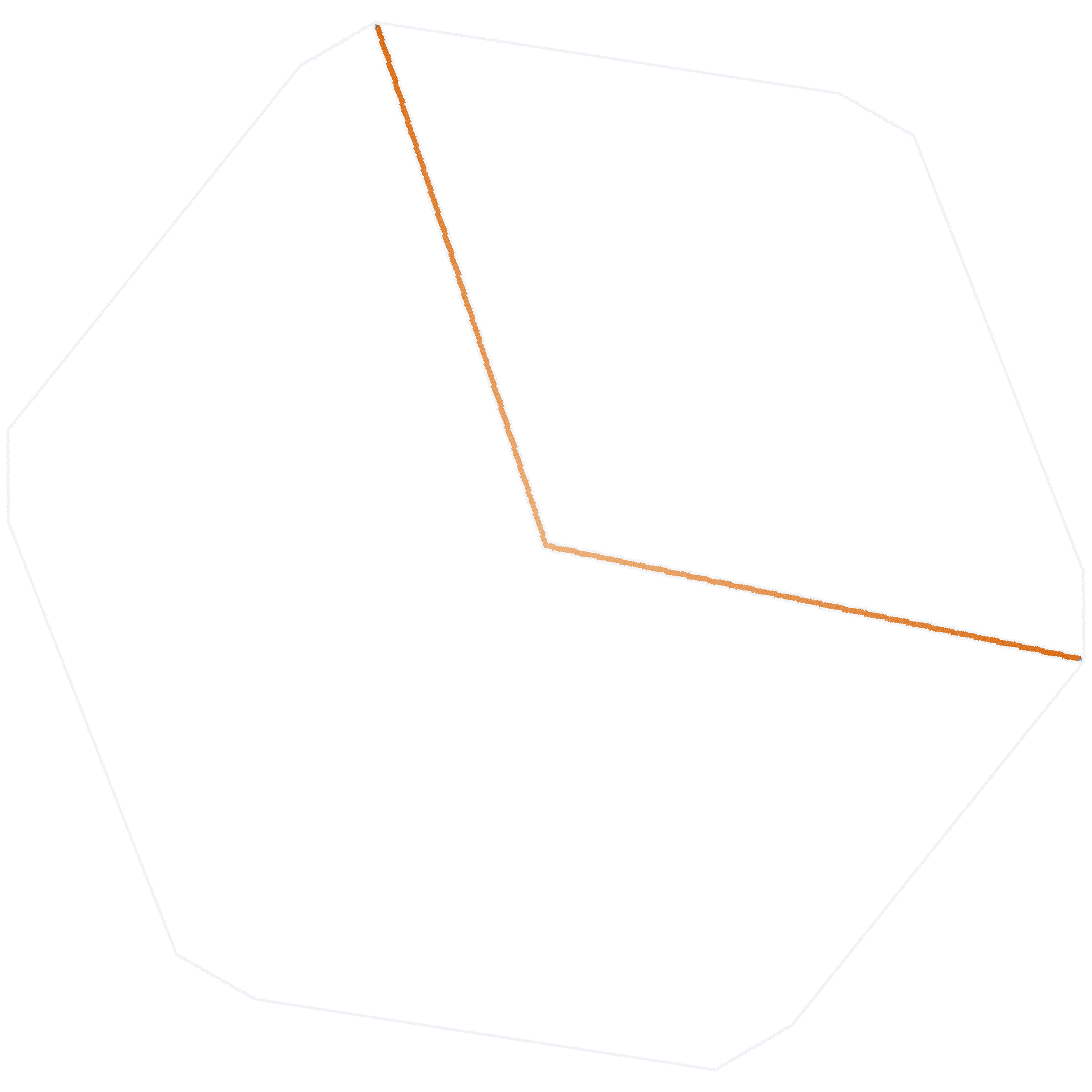}}
\centerline{\footnotesize (c)}
\caption{The spectre glider. (a) One glider of the pair at
generation 100 --- a single head cell and its wing wake. (b)
Worldtube at generation 150, following one glider. (c) Full lifetime
at radius 384: 761 generations, 3,435 tiles visited; the two ribbons
leave the seed on lanes \(120^\circ\) apart and both
reach the boundary.}
\label{fig:atlas-s33}
\end{figure}

\subsection{The spectre looper}

\begin{figure}
\centerline{\begin{minipage}[b]{11pc}\centerline{\includegraphics[width=11pc]{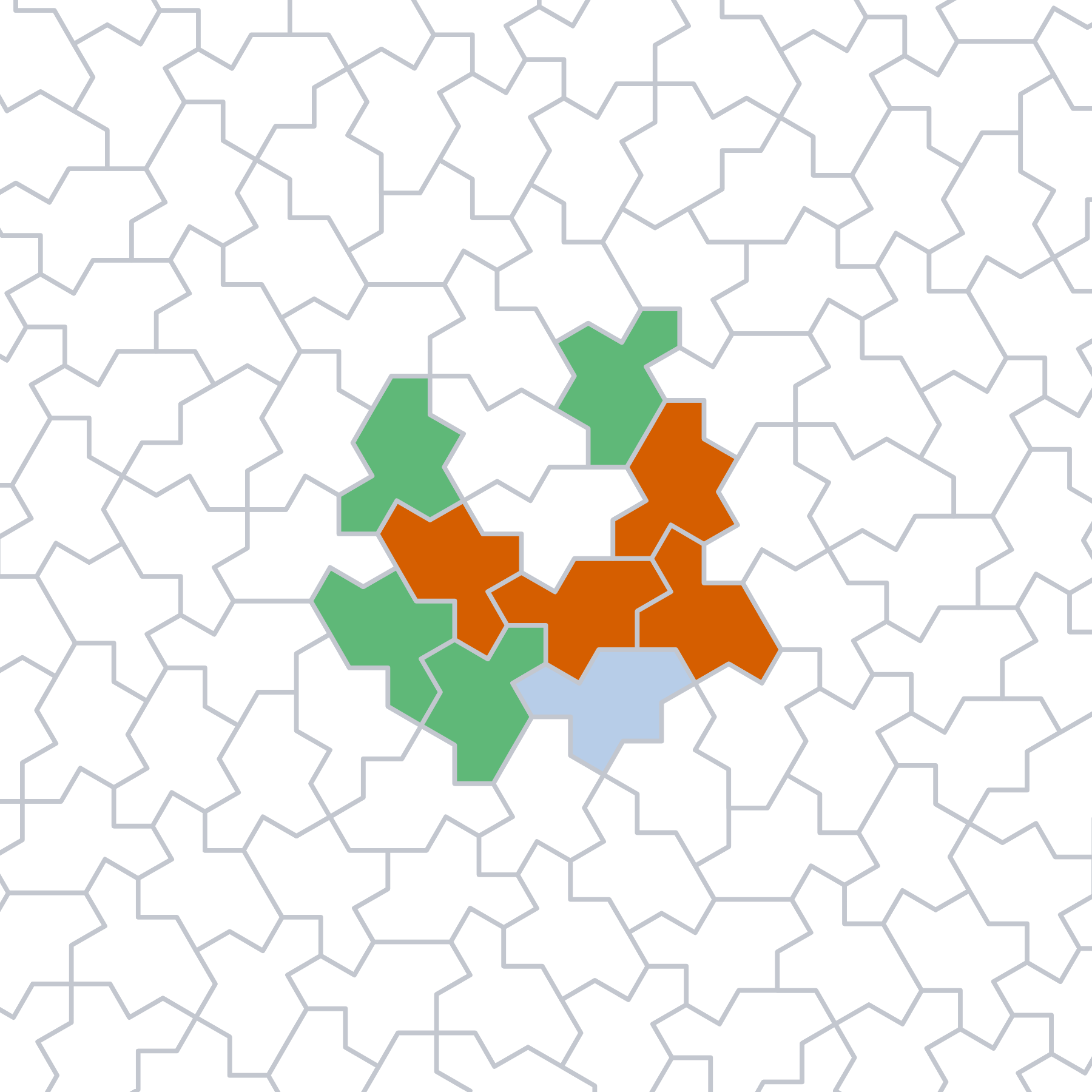}}\centerline{\footnotesize (a)}\end{minipage}\hskip 1pc
\begin{minipage}[b]{11pc}\centerline{\includegraphics[width=11pc]{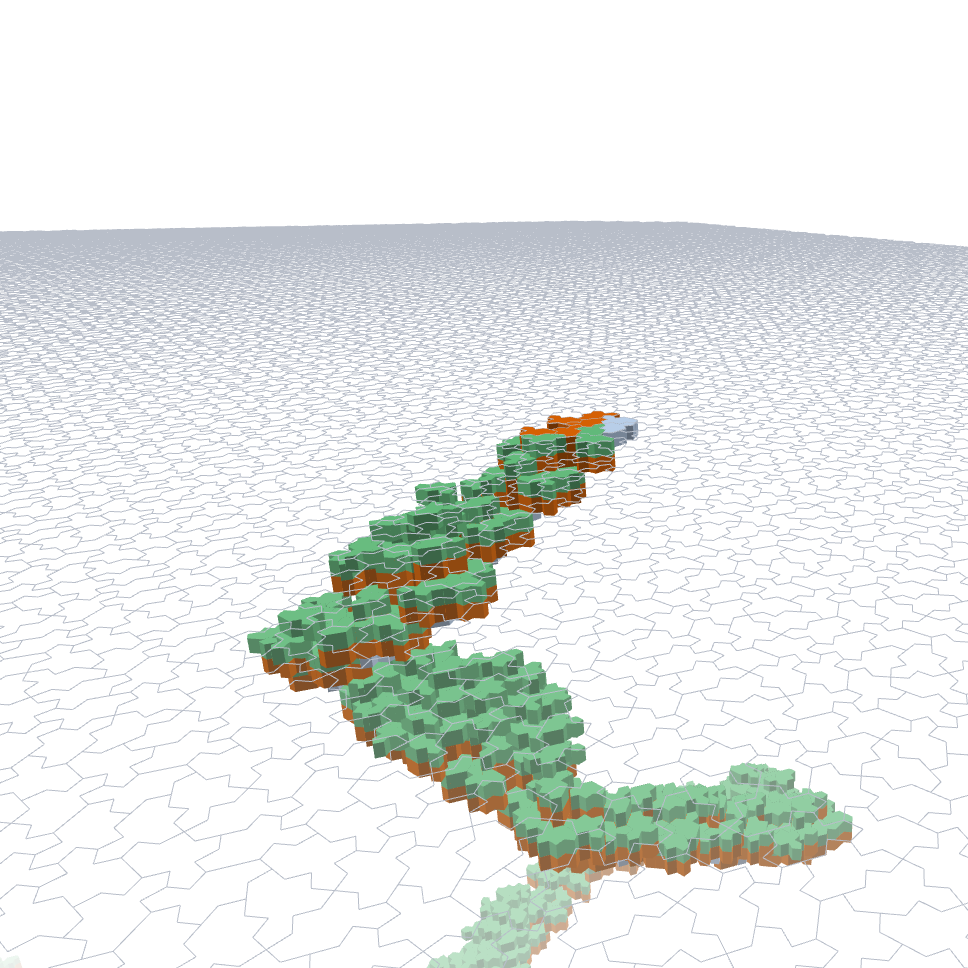}}\centerline{\footnotesize (b)}\end{minipage}}
\vskip 4pt
\centerline{\includegraphics[width=22pc]{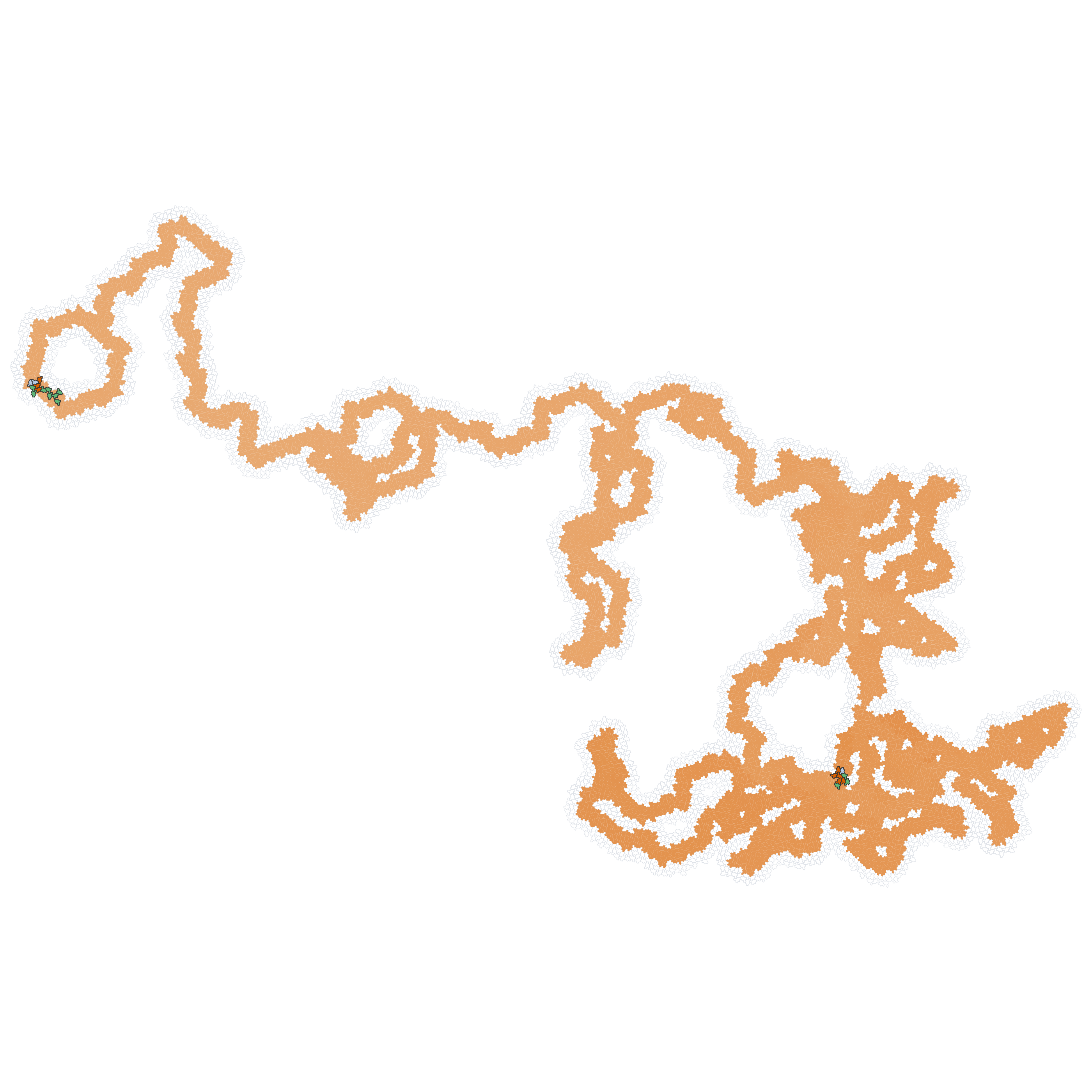}}
\centerline{\footnotesize (c)}
\caption{The spectre looper. (a) At generation 100 ---
indistinguishable from a glider at this scale. (b) Its worldtube
snakes and doubles back where a glider's climbs straight. (c) Full
lifetime over 4,800 generations at radius 384 (4,534 tiles visited),
zoomed to the trail: the pair separates at once; one wanderer is
trapped by generation 92 in the period-60 loop visible in the state
palette near the pale launch end, while its twin meanders for a
thousand generations before settling onto the period-550 circuit
threading the dense tangle at lower right. The patch boundary the
gliders reach lies far outside this frame.}
\label{fig:atlas-s32}
\end{figure}

Record \texttt{spectre-tableevolve-r48-s32} is the atlas's
counterexample --- the object the cross-radius standard rejected, and
the reason the standard matters. Its seed (3, 1) launches a pair of
oscillating wanderers with peak population 29--31; it reached full
fitness in its search and passed the radius-48 and radius-96 replays.
At radius 192 the classifier reports an exact recurrence of the full
patch state with period 3,300 --- the longest observed in this
project --- but that number is the beat of two independent captures,
not any single orbit. One wanderer is trapped almost at once: at
generation 92 it locks into a period-60 loop, a 207-tile track on
rings 17--40. Its twin wanders for a thousand generations, reaching
ring 189, before locking at generation 992 into a period-550 circuit
--- a closed track of 1,509 tiles spanning rings 107--189 --- and
from that moment the full state is periodic with period
\(\mathrm{lcm}(60, 550) = 3300\). At the anatomy scale
(Figure \ref{fig:atlas-s32}a) it is indistinguishable from a glider.
The lifetime views are what separate it: where every glider in this
atlas draws a straight ribbon into a corner of the patch, the
looper's trail (Figure \ref{fig:atlas-s32}c) wanders from its first
generation --- it never had the straight-lane signature --- and ends
in two closed orbits, the patch boundary far beyond the frame. Its worldtube snakes where the gliders' tubes climb straight.

\begin{table}
\centerline{\small\begin{tabular}{|c|c|c|}
\hline
Current state & Neighbor condition & Next state \\
\hline
0 & \(n_1 \geq 1\) and \(n_3 \geq 1\) & 3 \\
\hline
\inert{0} & \inert{\(n_1 \geq 1\) and \(n_3 \geq 1\)} & \inert{3} \\
\hline
0 & \(n_0 \geq 2\) and \(n_3 \geq 1\) & 1 \\
\hline
1 & \(n_0 \geq 2\) & 2 \\
\hline
0 & \(n_2 \geq 3\) & 2 \\
\hline
1 & always & 1 \\
\hline
\inert{3} & \inert{\(n_3 \geq 1\)} & \inert{1} \\
\hline
3 & \(n_0 \geq 1\) & 1 \\
\hline
any & otherwise & 0 \\
\hline
\end{tabular}}
\caption{The spectre looper's automaton. The grayed rows are inert:
row two duplicates row one, and every conversion row seven would
make, row eight makes identically.}
\label{tab:looperrule}
\end{table}

\section{Discussion}
\label{sec:discussion}

The contrast between the Penrose gliders and the monotile gliders
localizes what directed transport actually requires. Goucher's glider
is guided by a geometric rail whose parallelogram monotonicity also
supplies its immortality proof, and it holds its lane to within a
tile; the monotilings offer no rails, and their gliders wander in
corridors two to four tiles wide on the radius-384 trails while
holding heading to thousandths of a degree over a million rings ---
quantization equally sharp on both substrates. What the substrates share is not rails but
orientational order: a substitution tiling carries a globally
coherent orientation field, hierarchically propagated, with finitely
many tile orientations, and this compass evidently suffices to steer
a relay. On this reading the de Bruijn ribbons are the special case
in which the compass comes with a rigid lane, and the natural
conjecture is that any substitution tiling supports compass-locked
gliders once its rule space is expressive enough in the sense of the
ablation. The lanes lie on the measured fan exactly, and the fan is
the cusp structure of the graph metric --- the compass phase is
substrate geometry. Deriving that phase is now the open problem: the
cusps sit \(15.5^\circ\) (hat) and \(18.0^\circ\) (spectre)
from the tile-edge direction classes, on no crystallographic grid,
and their closed form plausibly lives in the tilings'
cut-and-project descriptions: each tiling has a model-set
representative in its conjugacy class, with pure-point diffraction
computed explicitly \cite{socolar, baakehat, baakespectre}.
One suggestive agreement deserves recording. The hat offset matches
\(180^\circ - 2\theta^*\), where
\(\theta^* = \cos^{-1}\!\bigl(\sqrt{2}/(4\phi^2)\bigr)\), with
\(\phi\) the golden ratio, is the twist angle between the two
hexagonal edge-direction families of the hat tiling's
inflation-invariant limit geometry \cite{socolar}:
\(45.5225^\circ\) against the measured
\(45.523^\circ \pm 0.003^\circ\), with no adjustable parameter. A
single-substrate match may yet be coincidence; the discriminating
test is the spectre, whose analogue must emerge from the
\(4 + \sqrt{15}\) arithmetic of its model set \cite{baakespectre}
with no further freedom. An exact computation of the
inflation-limit edge directions, in the frames the flights were
measured in, would decide it; we leave this to future work.

The taxonomy assembled across three substrates --- gliders,
wanderers, loopers --- reads less like three kinds of object than
one family in three regimes. Hat C's ignition shows a wanderer
decaying into three gliders, direct evidence that wanderers are
excited bound states of gliders; the Penrose wanderers travel in
fan-locked ballistic segments between stochastic reorientations; and
the loopers, from the
kite-and-dart orbits to the spectre looper, are glider
mechanisms whose paths close. The heading sweep even caught the
intermediate live: one launch ran briefly on one spoke, reoriented
once, and committed to the adjacent lane for a hundred thousand
rings --- a lane capture: a wanderer with exactly one segment
boundary. The shared relay clock and the shared
compass are the unifying measurements. In this picture the
interesting variable is not whether a rule supports motion but which
regime the motion lands in --- open lane, reorienting walk, or
closed orbit --- a classification reminiscent of the search-discovered
mobile objects of continuous cellular automata \cite{lenia}, here
with the substrate's geometry supplying the classifying structure.

Prospects for engineering must respect what aperiodicity removes. It
does not remove computation: a six-state semi-totalistic automaton
on the kite-and-dart tiling already simulates any boolean circuit
and any Turing machine \cite{imai}. What it removes is the economics
of construction. Constructive universality in the Game of Life rests
on components that, verified once, work everywhere, translation
symmetry making every site equivalent; the same guarantee underlies
glider-based universality in cellular automata generally
\cite{cook}. Here no symmetry carries one site to another: a
collision verified at one place holds only at the locally isomorphic
recurrences of that exact configuration, scattered at
substrate-chosen locations that must be found rather than reached by
translation. Engineering would therefore need site search plus
site-by-site verification, or components proven site-independent by
a certificate argument of the kind sketched below.

The observed repertoire is nonetheless a real starting kit:
reliable pair synthesis from two-cell seeds, annihilation by wake
contact, nondestructive deflection, and settlement onto
compass-commensurate mutual headings --- the last suggesting that
multi-glider configurations have a discrete, enumerable set of
stable geometries, which is exactly what an engineer wants. The
looper hints at the storage half of such a kit: it is the one
object here whose immortality is proved rather than extrapolated
(the exact cycle is a theorem about its patch), and its hexagonal
period-60 loop alone is, in principle, five bits of phase. But the
orbit is committed to the specific environment its wanderer happened
to find --- local isomorphism guarantees such
sites recur with bounded density, though not where --- and it inhabits a different rule from
the spectre glider, so the prerequisite for any signal-plus-storage
program is an object census within a single rule. An
alternative road avoids free flight altogether: signals confined to
constructed channels, in the manner of von Neumann's original
automaton \cite{vonneumann}, would trade the compass problem for a
plumbing problem
and deserve their own search.

The limitations are those already stated, plus scope. Immortality is
empirical, with the certificate problem open. The ablation's null
cells are bounded by budget --- two runs per family per cell ---
and assert difficulty, not emptiness. The rule space searched is
four-state semi-totalistic priority tables, a thin slice of local
rules. Discovery ran from a single root per family (ignition
genericity covers launch sites, not root choices). And every object
here was found by one fitness function; the searches were stopped by
success, not exhaustion, so the taxonomy is surely incomplete.

Future work is led by the certificate problem: can \textit{verified
to a million rings} be upgraded to \textit{immortal on every legal
tiling}? Substitution tilings have finite local complexity, and one
step of a cellular automaton is strictly local, so a proof reduces
to closure: enumerate the situations a flying glider occupies ---
its states together with the canonical collar graph, degrees
included --- and show the set closes under the step map, with
positive displacement, for every collar extension. Two structural
gifts make this tractable. Over-approximating the legal extensions
is safe (survival on a superset proves survival on the tiling), so
the substitution hierarchy is needed only where a lazy enumeration
fails; and the compass lane reduces the terrain ahead to a corridor
whose collar sequence is itself substitutive, making the certificate
a product of the glider's phase machine with a finite corridor
automaton --- the ribbon proof reappearing as the one-letter special
case.

We have run the decisive first experiment: the count of distinct
situations versus generation, with a cross-radius determinism
control validating the encoding (the same-collar successor map is
not forced even by a perfect encoding, so soundness is checked on
the forced maps from larger collars to smaller successors, all of
which are deterministic). The hat A glider occupies exactly 167
distinct situations at the step-sufficient collar radius. The last
new situation appears at generation 434; on the doubled patch, 1,074
consecutive further generations of entirely fresh terrain produce
nothing new. Branching --- situations with more than one observed
continuation --- is confined to four situations with two
continuations each, and their recurrence gaps carry the additive
structure of a substitutive sequence. The spectre glider gives the
same picture with a vocabulary of 306. Vocabularies are per-lane: the lane-capture trajectory
shares not one situation with the canonical glider's atlas, even
while visually indistinguishable from it, so a certificate covers
one rule--lane pair and six certificates cover a substrate. And
because semi-totalistic dynamics are blind to the tile classes the
keys record, the designated first step is a class-blind encoding: if
the six lanes' machines coincide there, as their identical clocks
and speeds suggest, each certificate factors into a small phase
machine crossed with a per-lane corridor automaton. The certificate
is therefore no longer a conjecture about whether a finite
description exists --- it measurably does --- but a bounded
enumeration task: generate the legal collar continuations of these
situations from the substitution hierarchy and verify closure of the
graph. Beyond the certificate: five and more
states and per-class stratified tables (is there a faster monotile
glider?); a deliberate collision study built on the interaction
repertoire; run-length statistics of wanderer segments, to locate
what sets the reorientation rate; and the
standing question of a Life-isomorphic automaton on the hat
\cite{baileylindsey}.

We close with a methodological remark. Nearly every substantive
claim in this project was, at some stage, almost false: the searches
invented four distinct ways to game their objectives, an early
symmetry claim died under a planned control, and the published table
we validated against contained a misprint. What made the surviving
results stable was not caution but structure --- pre-stated
kill-conditions, controls run before claims were recorded, and every
parasite promoted into a permanent filter. We offer the workflow,
in the tradition of open systematic exploration of simple programs
\cite{nks}, as a result alongside the results.

\textit{Note added.} After the first version of this paper was posted,
Andrew Trevorrow pointed out his \textit{HatLife} \cite{hatlife}: an
interactive environment, announced in May 2023, that runs
Generations-style and hand-designed rules on a fixed patch of 21,781
hats, with a spectre companion since. That version claimed that no
cellular automaton on either tiling had previously been reported. The
claim was wrong and is withdrawn. HatLife's exploration found
oscillators, a WireWorld implementation, and a four-state rule designed
so that chains called \textit{worms} travel the corridors between rows
of reflected hats, which the rule holds permanently empty; its body,
head, and tail states are all visible to neighbors, so it is a
priority-table rule in the sense of Section \ref{sec:methods} on the
vertex neighborhood, the one cell of the ablation of Section
\ref{sec:ablation} in which gliders were found. Those worms are
gliders in the sense of this paper. Transcribed into the present
machinery (vertex neighborhood; a two-stratum table rule, the
reflected hats given a table with no rows), worms arise from random
soups as Trevorrow reported, and two of them, of population eight and
seven, kept population and extent unchanged on patches of radius 48,
96, 192, and 384 while traveling, at two generations per ring, to the
boundary of each: the cross-radius criterion of Section
\ref{sec:methods}. Their corridors are not the rails of Section
\ref{sec:intro}: the reflected hats mark a direction, but nothing
confines a worm to it, and the worms wander within their corridors as
the gliders of this paper do. Their headings, measured at patch scale as for the
Penrose objects of Section \ref{sec:compass}, are \(225.524^\circ\)
and \(45.519^\circ\), on the hat compass of that section to a few
thousandths of a degree, the fit's standard error: an out-of-sample
test by a rule written by hand, for a different purpose, three years
before the compass was measured. The transcription and its record are
available from the author and will accompany a later release; they are
not part of the archived release of Section \ref{sec:repro}. What this
paper adds to that exploration is a searched rule space on either
tiling, objects followed beyond any fixed patch, and the compass law.

\section{Reproducibility}
\label{sec:repro}

Everything reported here regenerates from a small artifact set. The
source repository contains the engine, the search drivers, every
committed result record, the object compendium, and the figure
pipelines; each figure directory carries the script that regenerates
it from the records; the repository is public at
\href{https://github.com/jamesedmond/monotile-ca}{github.com/jamesedmond/monotile-ca}
and archived at Zenodo (doi:10.5281/zenodo.22836247).
A run is a record of tiling family, root address, radius,
neighborhood, rule, and seed, and every quantitative claim in this
paper maps to a record file and a single command; for example, the
radius-768 verification of hat glider A is one invocation of
the verifier against its committed record. The million-ring flights
regenerate the same way --- one invocation of the sliding-window
tracker against a committed seed record --- and the repository
carries each flight's decimated telemetry and its final checkpoint:
the glider at its millionth ring, independently replayable.

The companion interactive essay presents the same records live at
\href{https://offlattice.org/monotile/essay/}{offlattice.org/monotile/essay/}.
The engine that ran every search is compiled unchanged to
WebAssembly, so the browser replays are the evidence, not an
illustration of it: each panel replays a committed record, offers
the record itself for download, and displays the command that
re-verifies it.

Three determinism properties underwrite the scheme. Patches
regenerate exactly from their identifiers, with cell numbering
stable under radius growth, so a seed recorded at radius 48 lands on
the same tiles at radius 768. Searches are deterministic given their
recorded parameters and seeds. And the search code and the browser
run one code base, so verification and presentation cannot drift
apart.

\section*{Acknowledgments}
This work was carried out with substantial assistance from an AI
coding agent (Claude, Anthropic), which implemented the toolchain and
searches and drafted analyses under the author's direction; all
claims were verified by structural controls (pre-stated
kill-conditions, cross-radius and causality checks) and by the
author's independent review of the replayable records. Andrew Trevorrow
pointed out his 2023 HatLife and corrected the priority claim of an
earlier version of this paper.

\end{document}